\documentclass[reprint,superscriptaddress,amssymb,amsmath,aps,prd,longbibliography, nofootinbib]{revtex4-1}

\usepackage{graphicx}
\usepackage{amssymb}
\usepackage{epstopdf}
\usepackage{upgreek}
\usepackage{dcolumn}
\usepackage{bm}
\usepackage{gensymb}    
\usepackage{braket}
\usepackage{amsmath}
\usepackage{mathtools}
\usepackage{float}
\usepackage{tabularx}
\usepackage{mathrsfs}
\usepackage{lineno}
\usepackage{tikz}
\usepackage{graphicx}
\usepackage{sidecap} 
\usepackage{array}
\usepackage{amsthm}
\usepackage{enumitem}
\usepackage{xcolor} 

\expandafter\ifx\csname package@font\endcsname\relax\else
\expandafter\expandafter
\expandafter\usepackage
\expandafter\expandafter
\expandafter{\csname package@font\endcsname}%
\fi

\newcommand{\be}{\begin{eqnarray}}
\newcommand{\ee}{\end{eqnarray}}
\newcommand{\bfig}{\begin{figure}}
\newcommand{\efig}{\end{figure}}

\newcommand{\In}{\textsubscript{in}}
\newcommand{\Ina}{\textsubscript{in,$a$}}
\newcommand{\Inell}{\textsubscript{in,$\ell$}}
\newcommand{\Inm}{\textsubscript{in,$m$}}
\newcommand{\out}{\textsubscript{out}}
\newcommand{\dif}{\mathop{}\!\mathrm{d}}

\DeclareFontFamily{U}{mathb}{}
\DeclareFontShape{U}{mathb}{m}{n}{
  <-5.5> mathb5
  <5.5-6.5> mathb6
  <6.5-7.5> mathb7
  <7.5-8.5> mathb8
  <8.5-9.5> mathb9
  <9.5-11.5> mathb10
  <11.5-> mathbb12
}{}

\usepackage[colorlinks=true,allcolors=blue]{hyperref}%

\begin{document}

\title{Cavity entanglement and state swapping to accelerate \protect\\the search for axion dark matter}

\author{K. Wurtz}
\email{kwurtz@uwaterloo.ca}
\affiliation{Perimeter Institute for Theoretical Physics, 31 Caroline St.\ N., Waterloo, Ontario N2L 2Y5, Canada}
\affiliation{JILA, National Institute of Standards and Technology and the University of Colorado, Boulder, Colorado 80309, USA}

\author{B. M. Brubaker}
\affiliation{JILA, National Institute of Standards and Technology and the University of Colorado, Boulder, Colorado 80309, USA}
\affiliation{Department of Physics, University of Colorado, Boulder, Colorado 80309, USA}

\author{Y. Jiang}
\affiliation{JILA, National Institute of Standards and Technology and the University of Colorado, Boulder, Colorado 80309, USA}
\affiliation{Department of Physics, University of Colorado, Boulder, Colorado 80309, USA}

\author{E. P. Ruddy}
\affiliation{JILA, National Institute of Standards and Technology and the University of Colorado, Boulder, Colorado 80309, USA}
\affiliation{Department of Physics, University of Colorado, Boulder, Colorado 80309, USA}

\author{D. A. Palken}
\affiliation{JILA, National Institute of Standards and Technology and the University of Colorado, Boulder, Colorado 80309, USA}
\affiliation{Department of Physics, University of Colorado, Boulder, Colorado 80309, USA}

\author{K. W. Lehnert}
\affiliation{JILA, National Institute of Standards and Technology and the University of Colorado, Boulder, Colorado 80309, USA}
\affiliation{Department of Physics, University of Colorado, Boulder, Colorado 80309, USA}
\date{\today}

\begin{abstract}
In cavity-based axion dark matter detectors, quantum noise remains a primary barrier to achieving the scan rate necessary for a comprehensive search of axion parameter space. Here we introduce a method of scan rate enhancement in which an axion-sensitive cavity is coupled to an auxiliary resonant circuit through simultaneous two-mode squeezing (entangling) and state swapping interactions. We show analytically that when combined, these interactions can amplify an axion signal before it becomes polluted by vacuum noise introduced by measurement. This internal amplification yields a wider bandwidth of axion sensitivity, increasing the rate at which the detector can search through frequency space. With interaction rates predicted by circuit simulations of this system, we show that this technique can increase the scan rate up to 15-fold relative to the scan rate of a detector limited by vacuum noise.
\end{abstract}

\maketitle

\section{Introduction}
\label{sec:intro}

Several recent experiments in fundamental physics, including axion dark matter searches and gravitational wave searches, have reduced noise below the level of vacuum fluctuations \cite{backes2020quantum,LIGOsqueezing,megidish2019improved}. Not only do these searches require quantum-limited sensitivity, they must achieve it over a broadband frequency range. In the regime where quantum noise dominates, quantum measurement techniques hold unique potential to increase the bandwidth of these detectors.

Axion searches aim to detect a weak, narrowband signal whose frequency is \emph{a priori} unknown. This signal is generated by the axion field's hypothesized coupling to electromagnetism \cite{sikivie1985experimental}, through which dark matter axions in a magnetic field convert to photons with frequency corresponding to the axion's rest mass, $\omega\textsubscript{ax} = m\textsubscript{ax} c^2/\hbar$. Cavity-based axion detectors, known as haloscopes \cite{backes2020quantum, brubaker2017PRL, zhong2018results, du2018search, braine2020admx, lee2020CAPP}, are designed to generate and detect this signature using an electromagnetic cavity placed within a strong magnetic field. If $\omega\textsubscript{ax}$ is close to the resonance frequency of the cavity, the signal will cause a slight excess in the variance of the cavity's electric field \cite{malnou2019}.

Because the mass of the axion is unknown, the cavity is constructed to have a tunable resonance frequency. The resonance can then be adjusted in a step-wise manner, averaging noise for sufficient time to resolve an axion-induced excess at each tuning step. However, a comprehensive search of axion parameter space is still severely hindered by the achievable spectral scan rate of existing haloscopes. Even with noise reduced to the level of vacuum fluctuations, scanning the 1–10 GHz frequency band at the benchmark DFSZ coupling with modern haloscopes \cite{dfs1981,zhitnitsky1980} is estimated to take over 20,000 years of continuous detector live time \cite{palkenThesis}.

The scan rate is determined by two factors: the visibility, defined as the ratio of the power spectral density expected from an axion signal to the total noise power spectral density, and the visibility bandwidth, which is the characteristic bandwidth over which the detector remains sensitive. The visibility determines the required averaging time at each tuning step, while the visibility bandwidth determines the appropriate frequency step size. The linewidth $\Delta\textsubscript{ax}$ of the axion signal, determined by the velocity dispersion of the dark matter halo \cite{kenany2017}, is much smaller than the visibility bandwidth of a typical haloscope, so many distinct axion frequencies can be probed at each tuning step.

\begin{figure*}[t]
	\centering
	\includegraphics[scale=.99]{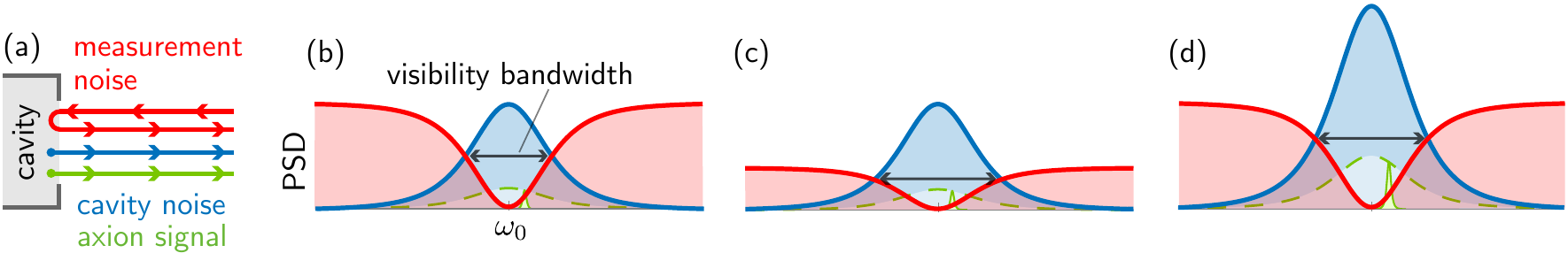}
	\caption{Noise in a single-quadrature for haloscopes with and without quantum noise manipulation.
	(a) In a haloscope detector, a weak axion signal (green) generated in the cavity emerges from the cavity, along with noise originating in the cavity (blue), and is then further polluted by measurement noise originating outside and reflecting off the cavity (red).
	(b) The axion signal, cavity noise, and measurement noise all contribute to the power spectral density near cavity resonance $\omega_0$, shown here in the absence of quantum noise manipulation. The narrow green peak indicates one possible manifestation of the axion signal, while the green dashed line indicates the power that would be delivered to the detector by an axion at a given detuning from $\omega_0$, which is proportional to the cavity noise PSD at that detuning. The double-headed arrow indicates the visibility bandwidth.
    (c) The measurement noise is prepared in a squeezed state. The axion and cavity noise contributions to the PSD are unchanged, but because the measurement noise dominates off resonance, squeezing widens the visibility bandwidth. (d) The axion signal and cavity noise are amplified together before being polluted by measurement noise, widening the visibility bandwidth. In each case, the cavity is assumed to be critically coupled (see Sec.\ \ref{sec:inputOutput}) for simplicity of representation.}
	\label{fig:intro}
\end{figure*}

Quantum-enhanced measurement techniques can widen the visibility bandwidthfa by increasing noise that originates in the cavity (along with any signal present) relative to noise associated with measurement. Figure \ref{fig:intro}(a) depicts the frequency dependence of cavity and measurement noise in the absence of quantum manipulation. The cavity noise, comprising thermal and vacuum fluctuations arising from the internal loss of the cavity by the fluctuation-dissipation theorem, is maximized on cavity resonance. The power expected from an axion signal would also be maximized if $\omega\textsubscript{ax}$ were to coincide with cavity resonance, and moreover, it exhibits the same roll-off as the cavity noise with increasing detuning of $\omega\textsubscript{ax}$ from resonance. If cavity noise were the only source of noise, then, the same visibility would be maintained on or off resonance. However, a second source of noise is introduced by the act of measurement itself. This noise, comprising thermal and vacuum fluctuations arising from loss external to the cavity, dominates off resonance and leads to a finite visibility bandwidth.\footnote{A third source of noise, contributed by the added noise of phase-insensitive amplifiers, has historically been an important contribution, but can be made negligible by performing phase-sensitive measurement \cite{backes2020quantum}.} 

One way to widen the visibility bandwidth is to suppress the amplitude of one phase of the measurement noise while amplifying the orthogonal phase \cite{malnou2019}, then measure only the suppressed (``squeezed'') component. These orthogonal phases (the cosine-like and sine-like components of the field) are called the $\hat{X}$ and $\hat{Y}$ quadratures. Preparing the measurement noise in a squeezed state, illustrated in Fig.\ \ref{fig:intro}(b), decreases the noise off resonance in the squeezed quadrature. This technique has been implemented experimentally \cite{backes2020quantum}, and results in a near-doubling of scan rate. However, it involves transporting the fragile single-mode squeezed state through lossy directional elements, which degrade the squeezing and limit the presently achievable scan rate enhancement.

In this article, we introduce a method to widen the visibility bandwidth by amplifying the cavity noise and axion signal together in a single quadrature relative to measurement noise, as illustrated in Fig.\ \ref{fig:intro}(c). This method involves using simultaneous two-mode squeezing (entangling) and state swapping interactions to realize a quantum non-demolition interaction between the cavity mode and an auxiliary resonant mode of a spatially separated readout circuit. Under this interaction, one quadrature of the cavity mode is mapped to the orthogonal quadrature of the auxiliary mode, and the measurement backaction is deposited in the unmonitored quadrature of the cavity mode. We will show that this method can yield more than a 15-fold scan enhancement relative to the quantum-limited haloscope scan rate. 

In the next section, we present a system capable of implementing this interaction, and show how the desired behavior arises from two-mode squeezing and state swapping interactions in a two-mode model of this system. In Sec.\ \ref{sec:inputOutput}, we apply an input-output theory analysis to this two-mode model to derive the scan rate enhancement as a function of the two-mode squeezing and state swapping interaction rates. In Sec.\ \ref{sec:transmissionLine}, we extend our analysis to account for effects outside the scope of the two-mode model. Finally, in Sec.\ \ref{sec:scanRate} we calculate the scan rate enhancement achievable with this system.

\begin{figure*}[t]
	\centering
	\includegraphics[scale=1]{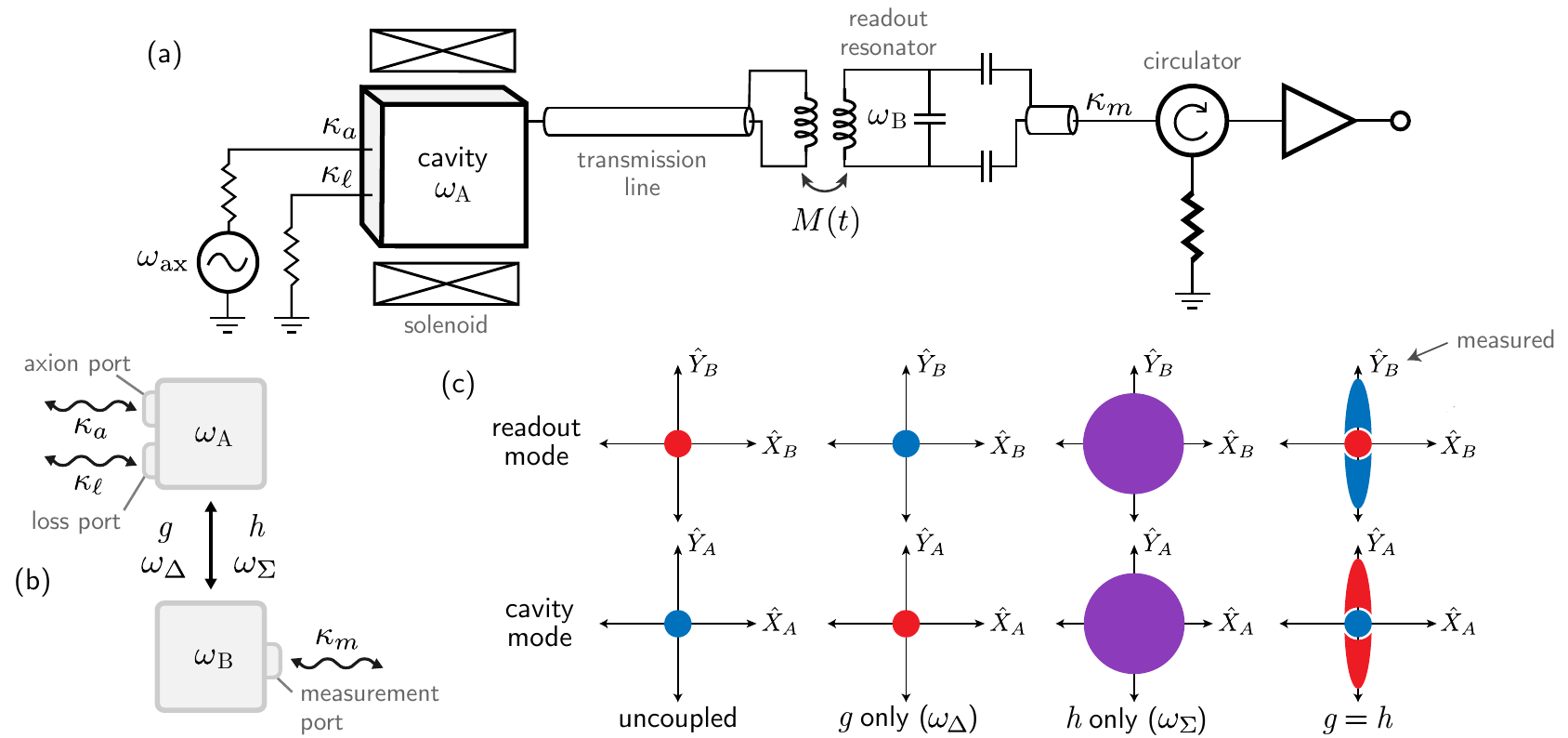}
	\caption{CEASEFIRE, a haloscope that uses state swapping and two-mode squeezing. (a) A microwave network model of CEASEFIRE. An axion-sensitive cavity with resonance frequency $\omega\textsubscript{A}$, axion field coupling rate $\kappa_a$, and loss rate $\kappa_\ell$ is coupled through a transmission line to an auxiliary readout resonator with frequency $\omega\textsubscript{B}$, which is coupled to a measurement chain at rate $\kappa_m$. The axion signal, modeled as a weakly coupled generator \cite{malnou2019}, and the cavity loss, modeled as a resistor, are coupled to the cavity by fictitious ports. The readout resonator can be modeled as a parallel $LC$ circuit coupled to the transmission line by a time-dependent mutual inductance $M(t)$; it is assumed to be superconducting such that its internal loss rate is negligible. A circulator shields the system from measurement chain backaction, but routes vacuum noise from a real resistor (bold) towards the measurement port. (b) A simplified two-mode model of CEASEFIRE. Modulating the mutual inductance simultaneously at the difference frequency $\omega_\Delta = \omega\textsubscript{B} - \omega\textsubscript{A}$ and the sum frequency $\omega_\Sigma = \omega\textsubscript{A} + \omega\textsubscript{B}$ creates state swapping and two-mode squeezing interactions at rates $g$ and $h$ respectively. (c) The uncoupled vacuum states of the readout resonator and cavity are represented by circles in quadrature phase space. Modulating $M(t)$ only at the difference frequency causes the cavity and readout modes to continuously swap states. Modulating only at the sum frequency entangles the states of the two modes, causing them to become amplified linear combinations of the uncoupled states. Modulating simultaneously at both frequencies and matching the interaction rates yields noiseless single-quadrature amplification. The measured readout mode quadrature contains measurement noise (reflected with unit magnitude) and amplified cavity noise, whereas the orthogonal quadrature contains only measurement noise.} 
	\label{fig:schematic}
\end{figure*}

\section{A haloscope using state swapping and two-mode squeezing}
\label{sec:implementation}

In order to overcome measurement noise, it would be ideal to amplify the axion signal and cavity noise together within the cavity. However, parametric amplifiers capable of noiseless amplification contain superconducting elements (Josephson junctions) incompatible with the strong magnetic field surrounding the cavity. Thus we consider a system in which the cavity is coupled through a transmission line to an auxiliary readout resonator that resides outside the magnetic field. We call this system CEASEFIRE --- the Cavity Entanglement And Swapping Experiment For Improved Readout Efficiency. A microwave network model of CEASEFIRE is shown in Fig.\ \ref{fig:schematic}(a). The readout resonator is modeled as a parallel $LC$ circuit coupled to the transmission line by a time-dependent mutual inductance $M(t)$ that is modulated by microwave-frequency drives. The time-dependent mutual inductance is an equivalent circuit model which captures the behavior of nonlinear devices such as the Josephson ring modulator (JRM) \cite{bergeal2010analog} and the tunable inductor bridge (TIB) \cite{chapman2016general}.

Fig.\ \ref{fig:schematic}(b) shows a two-mode model of the system when $M(t)$ is modulated at two distinct frequencies. Modulating at the difference of the mode frequencies $\omega_\Delta =\omega\textsubscript{B} - \omega\textsubscript{A}$ creates a beamsplitter interaction, in which the cavity mode ($\omega\textsubscript{A}$) and readout mode ($\omega\textsubscript{B}$) continuously swap states at rate $g$. This is illustrated in quadrature phase space in the second column of Fig.\ \ref{fig:schematic}(c), with cavity noise in blue and measurement noise in red. Modulating at the sum of the two frequencies $\omega_\Sigma = \omega\textsubscript{A} + \omega\textsubscript{B}$ prepares the cavity and readout modes in a two-mode squeezed state, wherein the quadrature variances and correlations between quadratures (a form of entanglement) grow exponentially at rate $h$. The final state of each mode in a two-mode squeezed state is an amplified linear combination of the initial states of the uncoupled modes, as illustrated in phase space in the third column of Fig.\ \ref{fig:schematic}(c). The interaction rates $g$ and $h$ are proportional to the amplitude of the modulation of $M(t)$ at frequencies $\omega_\Delta$ and $\omega_\Sigma$, respectively.

Simultaneously applying the state swapping and two-mode squeezing interactions and matching the interaction rates yields noiseless phase-sensitive amplification of the axion cavity mode. To illustrate this behavior, we first write the Hamiltonian of the two-mode system including parametric coupling interactions as
\begin{equation}
    \hat{H} = \hat{H}_0 + \hat{H}\textsubscript{int},
    \label{eq:Hamiltonian}
\end{equation}
where the uncoupled modes have the Hamiltonian
$\hat{H}_0 = \omega\textsubscript{A} \left( \hat{A}^\dag \hat{A} + \frac{1}{2} \right) + \omega\textsubscript{B} \left( \hat{B}^\dag \hat{B} + \frac{1}{2} \right)$ (with $\hbar = 1$). The interaction Hamiltonian is 
\begin{equation}
    \hat{H}\textsubscript{int} = \tilde{g} \hat{A} \hat{B}^\dag + \tilde{h} \hat{A}^\dag \hat{B}^\dag + \text{h.c.},
    \label{eq:interactionHamiltonian}
\end{equation}
where $\tilde{g} = g \exp(-i\omega_\Delta t)$,
$\tilde{h} = h \exp[-i(\omega_\Sigma t - \phi)]$, and $\phi$ is the phase difference between the microwave drives that generate the difference- and sum-frequency modulation.

By transforming to the quadrature basis and setting $h=g$, we obtain the quantum non-demolition interaction%
\begin{equation}
    \hat{H}\textsubscript{int} = 2 g \hat{X}_{\text{A},\phi/2} \hat{X}_{\text{B},\phi/2},
    \label{eq:HamiltonianQuadBasis}
\end{equation}
where $\hat{X}_{\text{A},\theta} = \frac{1}{\sqrt{2}} [e^{-i\theta}\hat{A} + e^{i\theta} \hat{A}^\dag]$ is the operator for a general cavity mode quadrature rotated by $\theta$ from $\hat{X}$, and $\hat{X}_{\text{B},\theta}$ is defined analogously. Setting $\phi = 0$ without loss of generality, the Heisenberg equation of motion for the $\hat{Y}$-quadrature of the readout mode is given by
\begin{equation}
    d\hat{Y}_{\text{B}}/dt = -2 g \hat{X}\textsubscript{A}.
\end{equation}
This expression indicates that the $\hat{X}$-quadrature of the cavity mode is mapped onto the $\hat{Y}$-quadrature of the readout mode, with noiseless gain that scales with $g$ \cite{hatridge2020,metelmannclerk2015}. 
On the other hand, we find that $d\hat{X}_{\text{B}}/dt = 0$, indicating that the orthogonal quadrature of the readout mode does not couple to the cavity mode at all. Thus, the combined interactions yield the behavior illustrated in the fourth column of Fig.\ \ref{fig:schematic}(c) --- noiseless single-quadrature amplification. 

\section{Two-mode CEASEFIRE model}
\label{sec:inputOutput}

To derive the visibility and scan rate enhancement of CEASEFIRE, we treat the two-mode model shown in Fig.\ \ref{fig:schematic}(b) in the input-output theory of quantum optics, in which loss and noise are modeled as arising from ports that couple propagating fields to the modes \cite{clerk2010review, walls2008quantum}. We use this framework to find the susceptibilities relating the output field at the measurement port to input fields at each port. To derive an expression for the scan rate enhancement, we then compare the signal visibility of CEASEFIRE to that of a haloscope without amplifier added noise whose total noise is subject to the quantum limit. We will refer to this latter system as a standard haloscope throughout the text.

As shown in Fig.\ \ref{fig:schematic}(a), the modes interact with the external environment via three ports parameterized by coupling rates. We model the axion field as a signal generator coupled to the cavity through a fictitious port with coupling rate $\kappa_a$ \cite{malnou2019}, and the cavity's internal loss as another fictitious port with coupling $\kappa_\ell$ equal to the rate at which power is dissipated internally. The coupling to the axion field is sufficiently weak that we assume $\kappa_a \ll \kappa_\ell$ in the analysis that follows. The readout resonator couples to a transmission line through a measurement port at rate $\kappa_m$, and is assumed to be superconducting such that its internal loss rate is negligible. The Heisenberg-Langevin equations of motion for $\hat{A}$ and $\hat{B}$, which include the terms associated with these coupling rates \cite{clerk2010review}, are given by
\begin{align}
    \cfrac{d\hat{A}}{dt} &= \begin{multlined}[t]
    -\Big( i \omega\textsubscript{A} + \cfrac{\kappa_\ell}{2} \Big) \hat{A}(t) + \sqrt{\kappa_\ell} \, \hat{\xi}\Inell(t) \\[10pt]
    + \sqrt{\kappa_a} \, \hat{\xi}\Ina(t) - i \tilde{g}^* \hat{B}(t) - i \tilde{h} \hat{B}^\dag(t)
    \raisetag{1.23cm}
    \label{eq:cavA_EqOfMotion}
    \end{multlined} \\
    \cfrac{d\hat{B}}{dt} &= \begin{multlined}[t]
    - \Big( i\omega_{\text{B}} + \cfrac{\kappa_m}{2} \Big) \hat{B}(t) + \sqrt{\kappa_m} \, \hat{\xi}\Inm(t) \\[10pt]
    - i \tilde{g} \hat{A}(t) - i \tilde{h} \hat{A}^\dag (t),
    \label{eq:cavB_EqOfMotion}
    \end{multlined}
\end{align}
where $\hat{\xi}\Inell$, $\hat{\xi}\Ina$, and $\hat{\xi}\Inm$ are the input fields incident on the loss, axion, and measurement ports, respectively. 

The ports' output fields are related to the input fields by 
\begin{equation}
    \hat{\xi}\textsubscript{out,$j$} = \hat{\xi}\textsubscript{in,$j$} - \sqrt{\kappa_j} \, \hat{\Xi},
    \label{eq:inputOutputRelations}
\end{equation} 
where $\hat{\Xi} = \hat{A}$ if $j = \{a, \ell \}$ and $\hat{\Xi} = \hat{B}$ if $j = m$. These equations of motion and input-output relations can be solved in the frequency domain to find the susceptibility matrix $\boldsymbol{\zeta}$, whose elements $\zeta_{jk}$ relate incoming fields at port $k$ to outgoing fields at port $j$. Transforming from the field operator basis to the quadrature basis, we can then derive expressions for the quadrature susceptibilities $\chi_{jk}$. The construction of the $\boldsymbol{\zeta}$ matrix and the derivation of the quadrature susceptibilities are outlined in Appx.\ \ref{appendix:inputoutput}. Here we present final expressions for three quadrature susceptibilities whose behavior is sufficient to describe the physics of CEASEFIRE. 

The response at the measurement port output, in the amplified quadrature, to a signal incident on the axion port, in the orthogonal quadrature, is given by
\begin{equation}
    \chi_{ma}(\omega) = \frac{(g + h)\sqrt{\kappa_m \kappa_a}}{g^2 - h^2 + (i\omega + \frac{\kappa_m}{2})(i\omega + \frac{\kappa_\ell}{2})},
    \label{eq:chi_ma}
\end{equation}
where $\omega$ is the detuning of the cavity input field from $\omega\textsubscript{A}$ (or equivalently, the detuning of the frequency-shifted readout resonator output field from $\omega_{\text{B}}$). The presence of $g$ and $h$ in the numerator reveals that the axion signal experiences gain. With $g = h$, the denominator becomes independent of $g$ and $h$; thus, the susceptibility bandwidth is independent of gain \cite{metelmannclerk2015}, and the gain can be increased by turning up $g$ and $h$ together. It is precisely the gain-independence of the susceptibility bandwidth that causes the \emph{visibility} bandwidth to increase with increasing gain.

The amplified-quadrature response at the measurement port due to noise in the orthogonal quadrature incident on the loss port is proportional to Eq.\ \eqref{eq:chi_ma}, scaled by the ratio of the port couplings:
\begin{equation}
    \chi_{m\ell}(\omega) = \sqrt{\frac{\kappa_\ell}{\kappa_a}} \chi_{ma}(\omega).
    \label{eq:chi_ml}
\end{equation}
The proportionality of these frequency-dependent susceptibilities shows that, as illustrated in Fig.\ \ref{fig:intro}(c), amplification does not improve the ratio of axion signal to cavity noise at any frequency. However, the gain experienced by the signal and cavity noise together is of primary importance, as it determines the visibility bandwidth. 

We quantify the gain of the axion signal and cavity noise by comparing the CEASEFIRE signal susceptibility to that of a standard haloscope \cite{malnou2019}, 
\begin{equation}
\chi_0(\omega) = \frac{\sqrt{\kappa_{c} \kappa_a}}{i\omega + (\kappa_{c} + \kappa_\ell)/2},
\end{equation}
with measurement port coupling rate $\kappa_c$. We define the gain as the ratio $|\chi_{ma}(\omega)/ \chi_0(\omega)|^2$, where the standard haloscope is taken to be critically coupled ($\kappa_c = \kappa_\ell$) with the same $\kappa_\ell$ and $\kappa_a$ as CEASEFIRE. On resonance and for $h = g$, the gain becomes
\begin{align}
    \bigg| \frac{\chi_{ma}(0)}{ \chi_0(0)} \bigg|_{h=g}^2 &= \frac{64 g^2}{\kappa_m \kappa_\ell} = 16 C,
    \label{eq:chi_ma_powerGain}
\end{align}
where we have introduced the cooperativity $C = 4 g^2/(\kappa_m \kappa_\ell)$, a dimensionless measure of interaction strength. With realistic values for these parameters (Sec.\ \ref{sec:scanRate}) we expect $C \approx 2500$ to be achievable on resonance, resulting in gain $\sim 4\times10^4$. Away from resonance, the gain rolls off as $\omega^{-2}$ for $\kappa_\ell \ll \omega \ll \kappa_m $ and $\omega^{-4}$ for $\omega \gg \kappa_m$.

The response at the measurement port output for a signal incident on the same port, i.e., the measurement port reflection susceptibility, is given by 
\begin{equation}
    \chi_{mm}(\omega) = 1 - \frac{\kappa_m ( i\omega + \frac{\kappa_\ell}{2} )}{g^2 - h^2 + (i\omega + \frac{\kappa_m}{2})(i\omega + \frac{\kappa_\ell}{2})}.
    \label{eq:chi_mm}
\end{equation}
Unlike the transmission susceptibilities, the reflection susceptibility is not phase-sensitive, so this relation holds for any quadrature. For $g = h$, the reflection susceptibility becomes $1 - \kappa_m/(i\omega + \frac{\kappa_m}{2})$, which has unit magnitude for all frequencies.

Next, we calculate the output spectral density at the measurement port in the presence of thermal noise and an axion signal. Thermal noise manifests as a white noise spectral density $n_T + 1/2$ incident on the loss and measurement ports, where $n_T = \text{exp}[(\hbar\omega/k_{\text{B}} T) - 1]^{-1}$ is the mean thermal photon occupancy of a mode with frequency $\omega$ in equilibrium with an environment at temperature $T$. Within a narrow frequency range $\lesssim \Delta\textsubscript{ax}$ around the axion signal frequency, the signal can be completely characterized by a spectral density $n_a \gg 1$, coupled at a rate $\kappa_a$ sufficiently small that $\kappa_a n_a \ll \kappa_\ell(n_T + 1/2)$. The relationship between the input-output theory model parameters $n_a$ and $\kappa_a$ and the parameters of the axion field is derived in Appx. B of Ref.~\cite{malnou2019}.
Specifically, we consider the amplified-quadrature measurement port output spectral density, denoted by $\mathbb{S}$. The contribution to $\mathbb{S}$ due to an axion signal is given by
\begin{equation}
    \mathbb{S}\textsubscript{$a$}(\omega) = n_a|\chi_{ma}(\omega)|^2 = \frac{\kappa_m \kappa_a n_a (g + h)^2}{|\beta(\omega)|^2},
    \label{eq:SoutmaAmped}
\end{equation}
where $\beta(\omega) = g^2 - h^2 + (i\omega + \kappa_m/2)(i\omega + \kappa_\ell/2)$. The contribution due to cavity noise and measurement noise is given by
\begin{equation}
    \mathbb{S}\textsubscript{$N$}(\omega) = \left( n_T + \frac{1}{2} \right) \left[ 1 + \frac{2 \kappa_\ell \kappa_m h (g + h) }{|\beta(\omega)|^2} \right].
    \label{eq:SoutmtotAmped}
\end{equation}
The visibility, which determines how long power must be averaged to resolve an axion-induced excess, is given by
\begin{equation}
    \alpha\textsubscript{CF}(\omega) = \frac{\mathbb{S}_a(\omega)}{\mathbb{S}_N(\omega) + \frac{1}{2}},
    \label{eq:visibilityCF}
\end{equation}
where the half quantum of noise added to the spectral density in the denominator accounts for the added noise of the subsequent quantum-limited phase-insensitive amplifier \cite{caves1982quantum}.\footnote{Because the gain decreases rapidly off resonance, a low-noise secondary amplifier is necessary to preserve the visibility bandwidth. In principle, we could use a single-quadrature amplifier for this purpose to avoid the half quantum of noise added by a phase-insensitive amplifier. However, this would require added operational complexity and would only confer a small benefit to overall scan rate enhancement.} We compare the amplified-quadrature axion visibility of CEASEFIRE to that of a standard haloscope:
\begin{align}
    \alpha_0(\omega) &= \frac{n_a |\chi_0(\omega)|^2}{n_T + \frac{1}{2}} \nonumber \\
   &= \frac{n_a \kappa_a \kappa_c}{\left(n_T + \frac{1}{2}\right) \left[ (\kappa_c + \kappa_\ell)^2/4 + \omega^2 \right]},
    \label{eq:visibilityQL}
\end{align}
where the limit $n_T \rightarrow 0$ (or $k_{\text{B}} T \ll \hbar \omega $) corresponds to the quantum-limited visibility. 

\begin{figure}[t]
	\centering
	\includegraphics[scale=1]{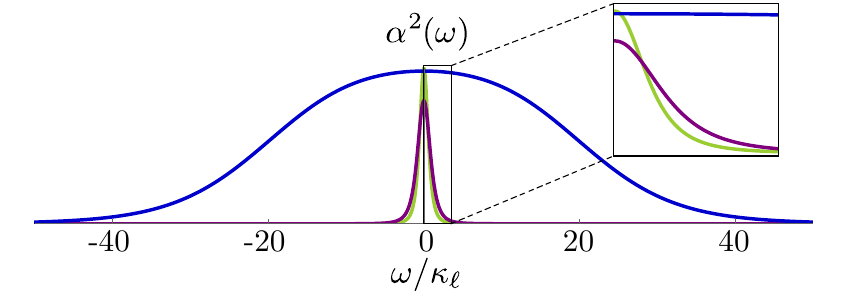} 
	\caption{Visibility improvement in the two-mode CEASEFIRE model. Plotting the squared visibility as a function of frequency shows that the two-mode CEASEFIRE model with $g = h = 110 \kappa_\ell$ and an optimal overcoupling of $\kappa_m = 19 \kappa_\ell$ (blue) greatly widens the bandwidth as compared to both a standard haloscope with $\kappa_c = \kappa_\ell$ (green), and a standard haloscope with $\kappa_c = 2\kappa_\ell$ (purple). Here, $\kappa_\ell$ is assumed to be the same in each case.} 
	\label{fig:2modeCEASEFIRE}
\end{figure}

The spectral scan rate scales as $\int \alpha^2(\omega) \dif \omega$. To define scan rate enhancement, we compare the CEASEFIRE scan rate to that of the twice-overcoupled standard haloscope ($\kappa_c = 2\kappa_\ell$), as a twofold overcoupling yields the maximum scan rate for a standard haloscope \cite{kenany2017}. In any haloscope experiment, overcoupling the cavity reduces visibility but increases bandwidth, and at some point an optimum that maximizes this integral is achieved. In the case of CEASEFIRE, amplification causes this tradeoff to be optimized at a much higher overcoupling --- strongly overcoupling widens the bandwidth, and operating at high cooperativity recovers peak visibility on resonance and over this wider bandwidth. In Fig.\ \ref{fig:2modeCEASEFIRE} we show the increase in visibility bandwidth of CEASEFIRE as compared to a standard haloscope.

The scan rate enhancement predicted by this two-mode input-output theory model is thus given by the ratio
\begin{equation}
    E = \frac{\int \dif \omega \, [\alpha\textsubscript{CF}(\omega)]^2}{\int \dif \omega \, [\alpha_0(\omega) ]^2_{\kappa_c = 2\kappa_\ell}},
    \label{eq:scanRateEnhancement}
\end{equation}
which is independent of the axion signal parameters $\kappa_a$ and $n_a$. Notably, the scan rate enhancement is also independent of $n_T$: the CEASEFIRE technique is equally beneficial for detectors at low frequencies or elevated temperatures.

\section{Physical implementation}
\label{sec:transmissionLine}

Our analysis thus far has considered the physics of CEASEFIRE in an idealized two-mode system. In this section, we discuss effects that would arise in an implementation of this concept. In particular, we will identify two effects we must mitigate to realize a significant scan rate enhancement: interactions between the readout mode and standing wave modes of the transmission line shown in Fig.\ \ref{fig:schematic}(a), and partial hybridization of the various modes in this multimode system. Here we present a qualitative overview of these effects; details of the extended model are relegated to Appx.\ \ref{appendix:transmissionline}.

As discussed in Sec.\ \ref{sec:implementation}, modulating the inductance of the readout resonator requires Josephson junctions, which are incompatible with the magnetic field around the cavity. Typical high-field magnet configurations can null the field approximately $L = 50\text{ cm}$ away from its maximum. Thus, the system requires a transmission line of this length separating the cavity and the readout resonator, and this transmission line introduces a spectrum of standing wave modes spaced by the free spectral range (FSR) $\Delta \textsubscript{FSR}/2\pi = v/(2L)$, where $v$ is the speed of light in the line.

In general, the normal modes of the system will be linear combinations of the uncoupled cavity mode, readout mode, and transmission line modes. Hybridization of the cavity mode with the readout mode creates an additional cavity mode loss channel, reducing the scan rate. We can eliminate most of this hybridization by using four equal inductors $L_0$ in a Wheatstone bridge configuration as the inductance of the readout circuit, and applying an inductance modulation $\pm M(t)$ to each element in the bridge with opposite sign for adjacent inductors (see Fig.\ \ref{fig:TL_circuit_model} in Appx.\ \ref{appendix:transmissionline}). This bridge coupler is electrically equivalent to mutual inductive coupling where the mutual inductance $M(t)$ can vary about zero, such that the first nonzero contribution to the static coupling between the cavity and readout modes arises at second-order in the fractional inductance modulation $M(t)/L_0$. The coupling inductance is modulated at both the difference frequency and the sum frequency of the cavity and readout modes to generate the dynamic couplings $g$ and $h$. 

Interactions between most other pairs of modes can be neglected: neither the sum nor the difference of their resonant frequencies is close to the modulation frequencies, so their effects on the equations of motion will vanish under a rotating wave approximation. However, the dynamic coupling between the readout mode and the two transmission line modes spectrally closest to the cavity mode can impact the scan rate enhancement significantly. Thus we extend our input-output theory analysis to a four-mode model, which we can then use to obtain predictions for the visibility and the scan rate enhancement given $\kappa_m$, $\kappa_\ell$, and the values of $g$ and $h$ obtained from a circuit model of the system.

The effect of these unwanted transmission line mode couplings is an effective mismatch in interaction rates that results in amplification of reflected measurement noise. Because CEASEFIRE widens the visibility bandwidth by amplifying cavity noise relative to measurement noise, this can reduce the scan rate enhancement considerably. When the cavity is spectrally centered between two transmission line modes, however, the effects of the transmission line mode interactions cancel almost entirely. In Sec.\ \ref{sec:scanRate}, we present results for the visibility and scan rate enhancement under the assumption that the cavity mode remains midway between the two nearest transmission line modes as it is tuned. In Appx.\ \ref{appendix:transmissionline} we discuss the dependence of the scan rate enhancement on detuning from this optimal operating point, as well as the scan rate enhancement that can be achieved by tuning transmission line modes.

\begin{figure}[b]
	\centering
	\includegraphics[scale=1]{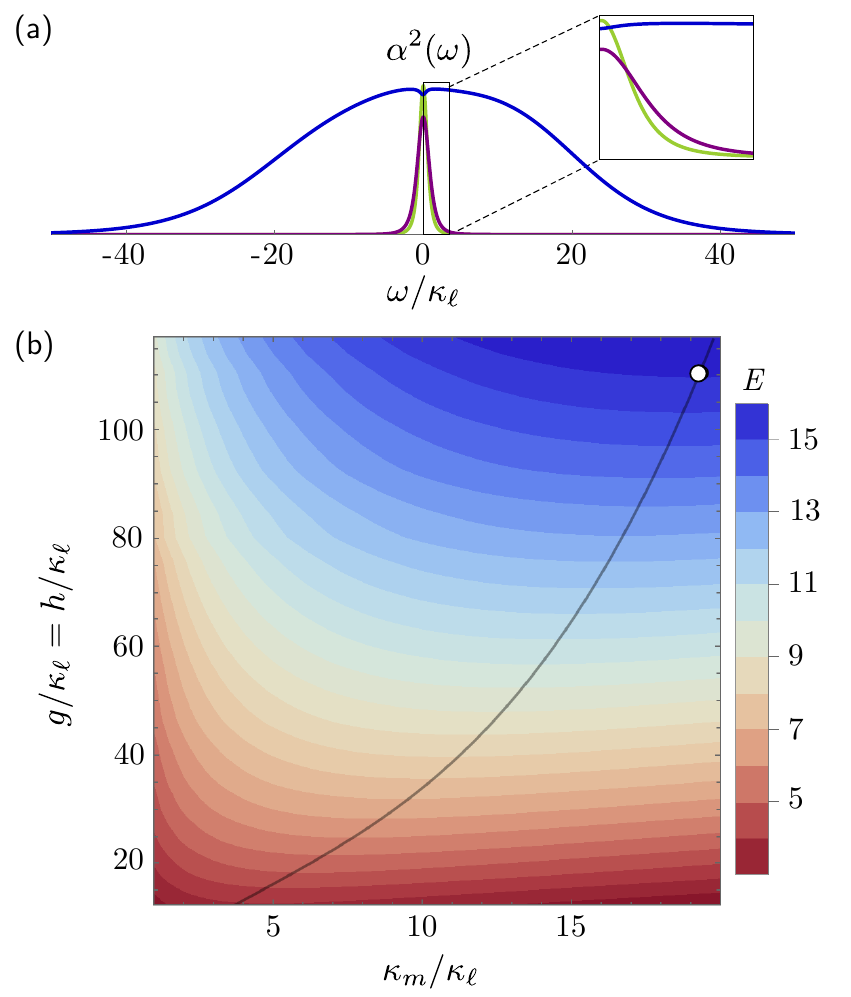} 
	\caption{Visibility and scan rate enhancement of the four-mode CEASEFIRE model. We take the cavity frequency to lie midway between two transmission line mode frequencies, optimizing the scan rate enhancement of the four-mode model. (a) Plotting with the same parameters and corresponding colors as in Fig.\ \ref{fig:2modeCEASEFIRE}, the four-mode model results in a qualitatively similar increase in visibility bandwidth as compared to a standard haloscope. The slight asymmetry in the CEASEFIRE visibility originates from unequal effects of the two transmission line modes spectrally closest to the cavity mode. (b) In color scale we plot the the scan rate enhancement of CEASEFIRE relative to the maximum scan rate for a standard haloscope (achieved for $\kappa_c = 2 \kappa_\ell$), as a function of matched dynamical coupling rates $g = h$ and overcoupling ratio $\kappa_m/\kappa_\ell$, assuming $\kappa_\ell/(2\pi) \approx 100$~kHz. For any given $g = h$, there exists an overcoupling ratio that yields the maximal scan rate enhancement (line); the parameters yielding the CEASEFIRE visibility shown in (a) are marked in (b) by the white circle.} 
	\label{fig:visibility}
\end{figure}

\section{Predicted scan rate enhancement}
\label{sec:scanRate}

Figure \ref{fig:visibility}(a) shows the squared visibility obtained from the four-mode model of CEASEFIRE, as well as the two standard haloscope cases plotted in Fig.\ \ref{fig:2modeCEASEFIRE}. For $\kappa_\ell/(2 \pi) \approx 100$~kHz, typical for copper haloscope cavities, the chosen values of $g$ and $h$ correspond to $23\%$ fractional inductance modulation. We bound the fractional modulation here to ensure that our model properly describes the higher order parametric processes that would arise from implementation with a TIB or JRM circuit (see Appx.\ \ref{subsection:interactionRates}). 

Figure \ref{fig:visibility}(b) shows the scan rate enhancement as a function of matched dynamic coupling rates $g = h$ and degree of overcoupling, assuming $\kappa_\ell/(2 \pi) \approx 100$~kHz, with the black line indicating the optimal overcoupling ratio. The parameters of Fig.\ \ref{fig:visibility}(a) are marked by a white dot, and correspond to a scan rate enhancement $E \approx 15.6$. The scan rate enhancement when rates are mismatched is discussed in Appx.\ \ref{appendix:ghMismatch}, where we find we can obtain a small increase in scan rate with a slight $g>h$ mismatch.

\section{Conclusion}
We have introduced a system capable of amplifying an axion signal through simultaneous two-mode squeezing and state swapping interactions, and shown that this system can significantly widen a haloscope's visibility bandwidth, yielding up to a 15-fold enhancement over the quantum-limited scan rate. In addition, we have introduced methods for mitigating problems caused by the periodic standing wave modes of a transmission line required to spatially separate the auxiliary readout resonator from the high magnetic field surrounding the axion-sensitive cavity. These methods may also be useful for efforts to incorporate superconducting qubit single-photon counters into haloscope detectors \cite{dixit2020searching}.

A successful experimental implementation of this concept would be a significant milestone in the application of quantum technology to the axion search. The method outlined here would also yield an equivalent scan rate enhancement in axion searches at lower frequencies in which the axion-sensitive resonator has substantial thermal occupation \cite{chaudhuri2019optimal} (this is also true of single-mode squeezing \cite{malnou2019}). In contrast, the quantum enhancement achievable with single-photon counting depends strongly on thermal occupancy \cite{lamoreaux2013analysis}. Because a comprehensive search of axion parameter space will likely require several technologies to act jointly to further increase sensitivity and bandwidth, we note that CEASEFIRE may also be compatible with simultaneously squeezing measurement noise and other methods of scan rate enhancement. 
 
\section*{Acknowledgements} 
\label{section:acknowledgements}

The authors thank Maxime Malnou, Nicholas Frattini, and Kyle Quinlan for helpful discussions. This document was prepared with support from the resources of the Fermi National Accelerator Laboratory (Fermilab), a U.S. Department of Energy, Office of Science, HEP User Facility. Fermilab is managed by Fermi Research Alliance, LLC (FRA), acting under Contract No. DE-AC02-07CH11359. Additionally this work was supported by Q-SEnSE: Quantum Systems through Entangled Science and Engineering (NSF QLCI Award OMA-2016244) and the NSF Physics Frontier Center at JILA (Grant No. PHY-1734006).

\appendix

\section{Two-mode input-output theory model}
\label{appendix:inputoutput}

In this appendix, we will use the equations of motion [Eqs.\ \eqref{eq:cavA_EqOfMotion} and \eqref{eq:cavB_EqOfMotion}] together with the input-output relations [Eq.\ \eqref{eq:inputOutputRelations}] to derive the full $6\times6$ susceptibility matrix $\boldsymbol{\zeta}$ in the field operator basis. We will then transform to the quadrature basis, derive the condition for the angle of the amplified quadrature, and use this to derive the amplified quadrature transmission susceptibilities [Eqs.\ \eqref{eq:chi_ma} and \eqref{eq:chi_ml}]. Finally, we will derive the amplified-quadrature output spectral densities at the measurement port, Eqs.\ \eqref{eq:SoutmaAmped} and \eqref{eq:SoutmtotAmped}.

In the rotating frame of the readout mode [$\hat{A}(t) \rightarrow \hat{A}(t) e^{-i\omega_{\text{B}} t}$, $\hat{B}(t) \rightarrow \hat{B}(t) e^{-i\omega_{\text{B}} t}$], the cavity equations of motion are given by 
\begin{align}
    \cfrac{d\hat{A}}{dt} &= \begin{multlined}[t]
    \left( i \omega_\Delta - \cfrac{\kappa_\ell}{2} \right) \hat{A}(t) + \sqrt{\kappa_\ell} \, \hat{\xi}\Inell(t) \\[10pt]
    + \sqrt{\kappa_a} \, \hat{\xi}\Ina(t) - i g \hat{B}(t) e^{i \omega_\Delta t} \\[5pt]
    - i h e^{i \phi} \hat{B}^\dag(t) e^{i \omega_\Delta t}
    \end{multlined}
    \label{eq:cavA_EqOfMotion_rotFrame} \\
    \cfrac{d\hat{B}}{dt} &= \begin{multlined}[t]
    - \cfrac{\kappa_m}{2} \hat{B}(t) + \sqrt{\kappa_m} \, \hat{\xi}\Inm(t) \\[10pt]
    - i g \hat{A}(t) e^{-i \omega_\Delta t} - i h e^{i \phi} \hat{A}^\dag(t) e^{i \omega_\Delta t}.
    \end{multlined}
    \label{eq:cavB_EqOfMotion_rotFrame}
\end{align}
Solving these equations in the frequency domain yields
\begin{align}
    \hat{A}(\omega) &=  \begin{multlined}[t]
    [i(\omega - \omega_\Delta) + \kappa_\ell/2]^{-1}[\sqrt{\kappa_\ell} \, \hat{\xi}\Inell(\omega)\\[10pt]
    + \sqrt{\kappa_a} \, \hat{\xi}\Ina(\omega) - i g \hat{B}(\omega - \omega_\Delta) \\[10pt] 
    - i h e^{i \phi} \hat{B}^\dag (-\omega + \omega_\Delta)]
    \end{multlined}
    \label{eq:cavA_EqOfMotion_rotFrame_fourier} \\[10pt]
    \hat{B}(\omega) &=  \begin{multlined}[t]
    [i\omega + \kappa_m/2]^{-1}[\sqrt{\kappa_m} \, \hat{\xi}\Inm(\omega) \\[10pt] 
    - i g \hat{A} (\omega + \omega_\Delta) - i h e^{i \phi} \hat{A}^\dag(-\omega + \omega_\Delta)].
    \raisetag{1.15cm}
    \label{eq:cavB_EqOfMotion_rotFrame_fourier}
    \end{multlined}
\end{align}
These frequency-domain equations can be uncoupled by substituting $\hat{B}(\omega - \omega_\Delta)$ and $\hat{B}^\dag (-\omega + \omega_\Delta)$ into Eq.\ \eqref{eq:cavA_EqOfMotion_rotFrame_fourier} and $\hat{A} (\omega + \omega_\Delta)$ and $\hat{A}^\dag(-\omega + \omega_\Delta)$ into Eq.\ \eqref{eq:cavB_EqOfMotion_rotFrame_fourier}, yielding $\hat{A}(\omega)$ and $\hat{B}(\omega)$ purely in terms of input fields:
\begin{widetext}
\begin{align}
    \hat{A}(\omega) &= 
    \begin{multlined}[t] 
    \beta(\omega - \omega_\Delta)^{-1} \Big\{ \Big[ i(\omega - \omega_\Delta) + \frac{\kappa_m}{2} \Big] \Big[ \sqrt{\kappa_\ell} \, \hat{\xi}\Inell(\omega) + \sqrt{\kappa_a} \, \hat{\xi}\Ina(\omega) \Big] - i g \sqrt{\kappa_m} \hat{\xi}\Inm(\omega - \omega_\Delta) \\[10pt]
    - i h e^{i \phi} \sqrt{\kappa_m} \hat{\xi}^\dag\Inm (-\omega + \omega_\Delta) \Big\}
    \end{multlined} \\
    \hat{B}(\omega) &= 
    \begin{multlined}[t]
    \beta(\omega)^{-1} \Big\{ \left( i\omega + \frac{\kappa_\ell}{2}\right) \sqrt{\kappa_m} \, \hat{\xi}\Inm(\omega) - i g [ \sqrt{\kappa_a} \, \hat{\xi}\Ina(\omega + \omega_\Delta) + \sqrt{\kappa_\ell} \, \hat{\xi}\Inell(\omega + \omega_\Delta) ] \\[10pt]
    - i h e^{i \phi} [ \sqrt{\kappa_a} \, \hat{\xi}^\dag\Ina(-\omega + \omega_\Delta) + \sqrt{\kappa_\ell} \, \hat{\xi}^\dag\Inell(-\omega + \omega_\Delta) ] \Big\}.
    \end{multlined}
\end{align}
\end{widetext}
The above equations and the general input-output relations given by Eq.\ \eqref{eq:inputOutputRelations} are then solved to obtain port susceptibilities. We define an input field vector 
\begin{multline}
    \vec{u} = [\hat{\xi}\Ina(\omega + \omega_\Delta),\: \hat{\xi}\Inell(\omega + \omega_\Delta),\: \hat{\xi}\Inm(\omega), \\ \hat{\xi}^\dag\Inm(-\omega),\: \hat{\xi}^\dag\Inell(-\omega + \omega_\Delta),\: \hat{\xi}^\dag\Ina(-\omega + \omega_\Delta)]^{\text{T}}
    \label{eq:inputfieldvector}
\end{multline} 
and an output field vector
\begin{multline}
    \vec{v} = [\hat{\xi}\textsubscript{out,$a$}(\omega + \omega_\Delta),\: \hat{\xi}\textsubscript{out,$\ell$}(\omega + \omega_\Delta),\:  \hat{\xi}\textsubscript{out,$m$}(\omega), \\
    \hat{\xi}^\dag\textsubscript{out,$m$}(-\omega),\: \hat{\xi}^\dag\textsubscript{out,$\ell$}(-\omega + \omega_\Delta),\: \hat{\xi}^\dag\textsubscript{out,$a$}(-\omega + \omega_\Delta)]^{\text{T}}
    \label{eq:outputfieldvector}
\end{multline}
such that we can write these relations compactly in the form $\vec{v} = \boldsymbol{\zeta} \vec{u}$, where $\boldsymbol{\zeta}$ is the susceptibility matrix. We can write the susceptibility matrix in the form
\[\boldsymbol{\zeta}(\omega) = 
\begin{bmatrix}
    \mathcal{S}(\omega) & \mathcal{P}(\omega) \\
    \mathcal{P}^\dag(-\omega) & \mathcal{S}^\ddag(-\omega)
\end{bmatrix},\]
where $\mathcal{S}$ is the symmetric matrix
\[\mathcal{S} = 
\begin{bmatrix}
     \zeta_{aa} & \zeta_{\ell a} & \zeta_{ma}  \\
    \zeta_{\ell a} & \zeta_{\ell\ell} & \zeta_{m \ell} \\ \zeta_{ma} & \zeta_{m \ell} & \zeta_{mm} \\
\end{bmatrix},\]
$\mathcal{P}$ is the persymmetric matrix 
\[\mathcal{P} = 
\begin{bmatrix}
    \zeta_{ma^\dag} & 0 & 0  \\
    \zeta_{m \ell^\dag} & 0 & 0 \\ 0 & \zeta_{m \ell^\dag} & \zeta_{ma^\dag}
\end{bmatrix},\]
$\mathcal{P}^\dag$ denotes the conjugate transpose of $\mathcal{P}$, and $\mathcal{S}^\ddag$ denotes the conjugate of $\mathcal{S}$ transposed across the anti-diagonal.

The individual susceptibility matrix elements are given by
\begin{align}
  \zeta_{jj}(\omega) &= 1 - \frac{\kappa_j (i\omega + \frac{\kappa_m}{2} )}{\beta(\omega)} \nonumber
  & \zeta_{mj}(\omega) &= \frac{i g \sqrt{\kappa_j \kappa_m} }{\beta(\omega)} \nonumber \\[5pt]
  \zeta_{mm}(\omega) &= 1 - \frac{\kappa_m (i\omega + \frac{\kappa_\ell}{2} )}{\beta(\omega)}   
  & \zeta_{mj^\dag}(\omega) &= \frac{i h e^{i\phi} \sqrt{\kappa_j \kappa_m} }{\beta(\omega)} \nonumber \\[5pt]
  \zeta_{\ell a}(\omega) &=  \frac{-\sqrt{\kappa_a \kappa_\ell} (i\omega + \frac{\kappa_m}{2} )}{\beta(\omega)}
  \label{eq:zetas}  
\end{align}
for $j = \{a,\ell\}$.

The central $2\times2$ submatrix of $\boldsymbol{\zeta}$, which describes reflection off the measurement port, is simply $\zeta_{mm}\mathbf{I}_2$, so it will remain unchanged under a unitary transformation to the quadrature basis. The quadrature reflection susceptibility is thus phase-independent, and given by $\chi_{mm} = \zeta_{mm}$, reproducing Eq.\ \eqref{eq:chi_mm} in the main text. The two amplified-quadrature transmission susceptibilities cited in the main text, Eqs.\ \eqref{eq:chi_ma} and \eqref{eq:chi_ml}, can be derived by first constructing the field in an arbitrary quadrature at the measurement port output, then identifying the phase condition for which the magnitude of the susceptibility is maximized.

The susceptibility matrix gives us the following expressions for the measurement port output fields due to an axion signal (omitting fields incident on other ports):
\begin{equation}
    \begin{split}
    \xi\textsubscript{out,$m$}(\omega) = 
    \frac{\sqrt{\kappa_a \kappa_m}}{\beta(\omega)} [ i g \hat{\xi}\Ina (\omega + \omega_\Delta ) \\
    + i h e^{i \phi} \hat{\xi}^\dag\Ina (-\omega + \omega_\Delta)] \vspace{1cm}
    \end{split}
\end{equation}
\begin{equation}
    \begin{split}
        \xi^\dag\textsubscript{out,$m$}(-\omega) = \frac{\sqrt{\kappa_a \kappa_m}}{\beta(\omega)} [ - i h e^{-i\phi} \hat{\xi}\Ina (\omega + \omega_\Delta) \\
    - i g \hat{\xi}^\dag\Ina (-\omega + \omega_\Delta) ].
    \end{split}
\end{equation}
\noindent To express these output fields in the quadrature basis, we substitute the above expressions into the general relation between field operators and an arbitrary quadrature operator rotated by $\theta$ from the $\hat{Y}$-quadrature:
\begin{equation}
    \begin{split}
        &\hat{Y}\textsubscript{out,$m,\theta$} (\omega) =
        \frac{1}{\sqrt{2}i} \big[ e^{-i\theta} \hat{\xi}\textsubscript{out,$m$}(\omega) - e^{i\theta} \hat{\xi}^\dag\textsubscript{out,$m$}(-\omega) \big] \\[3mm]
        & = \frac{\sqrt{\kappa_a \kappa_m/2}}{ \beta(\omega)} \, \times \\[2mm]
        &\hspace{2.5mm}\big\{ e^{-i\phi/2} \, \hat{\xi}\Ina (\omega + \omega_\Delta) \big[ g e^{i(\phi/2 - \theta)} + h e^{-i(\phi/2 - \theta)} \big] \\[2mm]
        &+ e^{i\phi/2} \, \hat{\xi}^\dag\Ina (-\omega + \omega_\Delta) \big[ g e^{-i(\phi/2 - \theta)} 
        + h e^{i(\phi/2 - \theta)} \big] \big\}.
        \label{eq:Youtmtheta}
        \raisetag{62pt}
    \end{split}
\end{equation}
Note that in the second equality above, for $\theta = \phi/2$, the terms inside the square brackets become independent of $\phi$. These $[g + h]$ terms can then be factored out, and the remaining expression inside the curly braces becomes that of a particular input quadrature, namely $\hat{X}\textsubscript{in,$a,\phi/2$}$. Thus, $\theta = \phi/2$ defines the amplified quadrature, in agreement with Eq.\ \eqref{eq:HamiltonianQuadBasis}.\footnote{For $\theta = \phi/2 + \pi/2$, the terms in square brackets become $\pm i(g - h)$. This shows that when $g = h$ the input quadrature orthogonal to the amplified quadrature is completely decoupled from the output.} The factor relating the amplified output quadrature $\hat{Y}\textsubscript{out,$m,\phi/2$} \equiv \hat{\mathbb{Y}}\textsubscript{out,$m$}$ to the corresponding input quadrature $\hat{X}\textsubscript{in,$a,\phi/2$} \equiv \hat{\mathbb{X}}\textsubscript{in,$a$}$ is then
\begin{equation}
   \chi_{ma}(\omega) = \frac{(g + h) \sqrt{\kappa_a \kappa_m}}{\beta(\omega)},
\end{equation}
\noindent given by Eq.\ \eqref{eq:chi_ma} in the main text. The loss port susceptibility $\chi_{m\ell}(\omega)$ [Eq.\ \eqref{eq:chi_ml}] is derived analogously. 

These susceptibilities and the noise of the input fields determine the noise of the output fields. The input noise in the field operator basis is characterized by the input spectral density matrix 
\begin{align}
    \textbf{S}\In &= \langle[\vec{u}(\omega)]^\dag \vec{u}^T(\omega)\rangle \nonumber \\
    &= \text{diag}[n_a, n_T, n_T, n_T + 1,n_T + 1,n_a + 1],
    \label{eq:inputSpectralDensityMatrix}
\end{align}
where the Hermitian conjugate in the first equality does not transpose the vector \cite{zheng2016accelerating}. The output spectral density matrix in the field operator basis is then given by $\textbf{S}\out(\omega) = \boldsymbol{\zeta}^*(\omega)\textbf{S}\In\boldsymbol{\zeta}^T(\omega)$.

We now calculate the output spectral density at the measurement port in the $\hat{Y}_\theta$ quadrature. The spectral density is given by the variance of this quadrature operator as defined in the first line of Eq.\ \eqref{eq:Youtmtheta}:

\begin{widetext} 
\vspace{-.15in}
    \begin{multline}
    S\textsubscript{out,$m,\theta$}(\omega) = \left\langle (\hat{Y}\textsubscript{out,$m,\theta$} (\omega))^2 \right\rangle = \frac{1}{2} \left[ \left\langle \hat{\xi}^\dag\textsubscript{out,$m$}(-\omega) \hat{\xi}\textsubscript{out,$m$}(\omega) \right\rangle + \left\langle \hat{\xi}\textsubscript{out,$m$}(\omega) \hat{\xi}^\dag\textsubscript{out,$m$}(-\omega) \right\rangle \right] - \frac{1}{2} \Big[ \left\langle (\hat{\xi}\textsubscript{out,$m$}(\omega))^2 \right\rangle \\[10pt]
    + \left\langle (\hat{\xi}^\dag\textsubscript{out,$m$}(-\omega))^2 \right\rangle \Big] (\cos^2\theta - \sin^2\theta) + i \left[ \left\langle (\hat{\xi}\textsubscript{out,$m$}(\omega))^2 - (\hat{\xi}^\dag\textsubscript{out,$m$}(-\omega))^2 \right\rangle \right] \cos\theta \sin\theta .\hspace{10mm}
    \raisetag{2cm} 
    \label{eq:outputSpectralDensityGeneralTheta}
    \end{multline}
\end{widetext}
The expectation values of field operator products that appear in Eq.\ \eqref{eq:outputSpectralDensityGeneralTheta} can be associated with elements of the output spectral density matrix through the relation $\textbf{S}\textsubscript{out}(\omega) = \langle[\vec{v}(\omega)]^\dag \vec{v}^\text{T}(\omega)\rangle$. 

Substituting in these elements, the contribution to this measurement port output spectral density from an axion signal is obtained from
\begin{align}
     S\textsubscript{out,$m,a,\theta$}(\omega) &= S\textsubscript{out,$m,\theta$}(\omega) - S\textsubscript{out,$m,\theta$}(\omega)\Big|_{n_a = 0} \nonumber \\
     & = \frac{n_a \kappa_a \kappa_m [g^2 + h^2 + 2 g h\text{cos}(2\theta - \phi)] }{|\beta(\omega)|^2},
     \label{eq:SoutmaTheta}
\end{align}
which is maximized for $\theta = \phi/2$, in agreement with the amplified quadrature condition derived from the susceptibility matrix. The amplified-quadrature axion signal spectral density at the measurement port is $S\textsubscript{out,$m,a,\phi/2$} \equiv \mathbb{S}_a$, given by Eq.\ \eqref{eq:SoutmaAmped}.

The output spectral density due to thermal and vacuum noise in the $\hat{Y}_\theta$ quadrature is given by
\begin{multline}
    S\textsubscript{out,$m,N,\theta$}(\omega) = S\textsubscript{out,$m,\theta$}(\omega)\Big|_{n_a = 0} \\
    = \left(n_T + \frac{1}{2}\right) \left[ 1 + \frac{2 \kappa_\ell \kappa_m (h^2 + g h\text{cos}(2\theta - \phi)) }{|\beta(\omega)|^2} \right],
    \raisetag{1.5cm}
    \label{eq:SoutmtotTheta}
\end{multline}
which, in the amplified quadrature, is $S\textsubscript{out,$m,N,\phi/2$} \equiv \mathbb{S}_N$, given by Eq.\ \eqref{eq:SoutmtotAmped}.

\section{Mismatching interaction rates}
\label{appendix:ghMismatch}

Thus far, we have asserted that CEASEFIRE behaves optimally when the swap rate $g$ is matched with the two-mode squeezing rate $h$. Here, we analyze the behavior of the two-mode model (presented in Sec.\ \ref{sec:inputOutput} and Appx.\ \ref{appendix:inputoutput}) when the rates are mismatched. The behavior of the extended model of Sec.\ \ref{sec:transmissionLine} and Appx.\ \ref{appendix:transmissionline} is qualitatively similar. Specifically, we determine how sensitive the measurement port reflection susceptibility and the scan rate enhancement are to small deviations from $g = h$. We will find that although the reflection susceptibility is highly sensitive to slight mismatch when operating at high cooperativity, the scan rate enhancement is quite insensitive to these changes, and is in fact optimized when $g$ is slightly larger than $h$.

The measurement port reflection susceptibility $\chi_{mm}(\omega)$, introduced in Eq.\ \eqref{eq:chi_mm}, is a useful quantity for understanding the behavior of CEASEFIRE with mismatched interaction rates. It is also of practical relevance, as it is directly accessible to the experimentalist. For our purposes it will be sufficient to consider $|\chi_{mm}(0)|$. As discussed in the main text, for $g = h$, the effects of the two interactions on the reflection susceptibility cancel, and the reflection susceptibility reduces to that of a single-port resonator, with unit magnitude for all frequencies.

We first consider only the state swap interaction, with the two-mode squeezing interaction turned off ($h = 0$). In the absence of any port decay rates, the modes would continuously swap states at rate $g$ (second column of Fig. \ \ref{fig:schematic}(c)). When ports are added to the model, the steady-state behavior depends on the value of $\kappa_m$ relative to the effective cavity loss rate as seen by fields incident on the measurement port, $\kappa_\text{eff}$, which depends only on $\kappa_\ell$ and $g$. In particular, for $\kappa_m = \kappa_\text{eff}$, the system will appear critically coupled, and all energy incident on the measurement port on resonance will be delivered to the loss port. Solving for $|\chi_{mm}(0)| = 0$, we find that $\kappa_\text{eff} = 4 g^2/ \kappa_\ell$. Defining the state swap cooperativity as
\begin{equation}
    C_g = \frac{4g^2}{\kappa_m \kappa_\ell},
\end{equation}
the critical coupling condition corresponds to $C_g = 1$.

Next, we consider only the two-mode squeezing interaction, with the state swapping interaction turned off ($g=0)$. In the absence of port decay rates, the cavity and readout mode quadrature variances and covariances grow exponentially at rate $h$ (third column of Fig. \ \ref{fig:schematic}(c)). When ports are added to the model, the system behaves as a two-mode amplifier that operates in reflection, with a fixed gain-bandwidth product centered on cavity resonance. The peak gain is determined by the two-mode squeezing cooperativity 
\begin{equation}
   C_h = \frac{4h^2}{\kappa_m \kappa_\ell}, 
\end{equation}
and formally diverges $(|\chi_{mm}(0)| \rightarrow \infty)$ for $C_h=1$. In a real system, this divergent gain would be capped by nonlinearities not included in our model, resulting in self-sustained oscillations with fixed amplitude for $C_h \geq 1$.

\begin{figure}[t]
	\centering
	\includegraphics[scale=1]{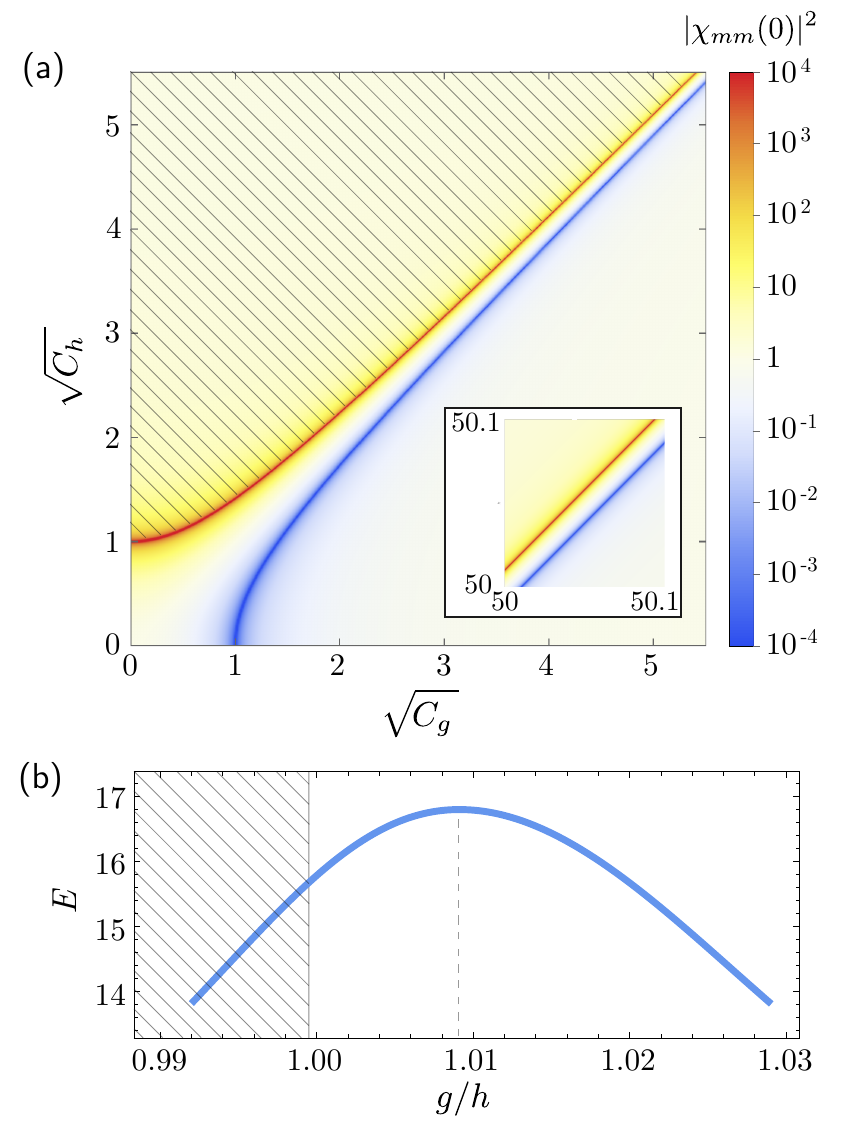}
	\caption{CEASEFIRE behavior with mismatched interaction rates. Panel (a) shows the squared reflection susceptibility on resonance as a function of state swap and two-mode squeezing cooperativities. We plot in terms of the square root of these quantities so as to show behavior linear in $g$ and $h$. In the hatched region of this parameter space, nonlinearities not included in our model of CEASEFIRE would result in self-sustained oscillations. The inset plot shows the increased sensitivity to mismatch at $C = 2500$, approximately the cooperativity corresponding to the operating parameters in Fig.\ \ref{fig:visibility}(a). Panel (b) shows the scan rate enhancement of the extended CEASEFIRE model for small changes around the values assumed in Fig.\ \ref{fig:visibility}(a). The hatched region corresponds to the hatched region in (a).} 
	\label{fig:ghmismatch}
\end{figure}

Fig.\ \ref{fig:ghmismatch}(a) shows $|\chi_{mm}(0)|^2$ as a function of the state swap and two-mode squeezing cooperativities (plotted in terms of $\sqrt{C_g}$ and $\sqrt{C_h}$ so as to plot linearly in $g$ and $h$). The behavior along the horizontal and vertical axes illustrates the pure state swapping and pure two-mode squeezing regimes. We see that with both interactions present, for any value of $g$, there is always some $h < g$ for which the system behaves as if critically coupled, and some $h > g$ for which it exhibits divergent gain. Moreover, the regions of critical coupling and divergent gain converge towards $g = h$ --- the system is much more sensitive to mismatch at high cooperativity.

To understand the increased sensitivity to mismatch at high cooperativity, we consider operation with interaction rates $g = g_0 + \delta g$ and $h = h_0 + \delta h$, with $g_0 = h_0$ and $\delta g, \delta h \ll g_0$ such that $g + h \approx 2g_0$. Then Eq.\ \eqref{eq:chi_mm} evaluated at $\omega = 0$ becomes
\begin{equation}
    \chi_{mm}(0) = 1 - \frac{\kappa_m \kappa_\ell/2}{2g_0(g-h) + \kappa_m \kappa_\ell/4}.
\end{equation}
We can rewrite this as
\begin{equation}
    \chi_{mm}(0) = 1 - \frac{2}{2C\varepsilon + 1},
\end{equation}
where $C$ is the cooperativity of the matched rates as defined in Sec.\ \ref{sec:inputOutput}, and $\varepsilon$ is the fractional rate mismatch given by $\varepsilon = (g - h)/g_0$. For $2C \varepsilon \rightarrow +1$, $|\chi_{mm}(0)|^2 \rightarrow 0$, and the system is critically coupled. For $2C \varepsilon \rightarrow -1$, $|\chi_{mm}(0)|^2 \rightarrow \infty$, and the system yields divergent gain. To remain close to $|\chi_{mm}(0)| = 1$ around $g = h$ we would require $|\varepsilon| \ll 1/(2C)$. For $C \approx 2500$, where we would expect an experiment to operate, this is a very stringent constraint.

Fig.\ \ref{fig:ghmismatch}(b) shows the scan rate enhancement $E$ of the extended CEASEFIRE model for small deviations around $g/h=1$, taking $h =  110\kappa_\ell$ and $\kappa_m = 19 \kappa_\ell$, as in Fig.\ \ref{fig:visibility}(a). This illustrates that a rate mismatch of $\sim 1\%$ has little effect --- the scan rate enhancement is much more tolerant to mismatch than the reflection susceptibility. To see this analytically, consider the visibility on resonance [Eq.\ \eqref{eq:visibilityCF}], setting $n_T = 0$ for simplicity:
\begin{align}
    \alpha\textsubscript{CF}(0) &= \frac{2\kappa_m \kappa_a n_a (g + h)^2}{(g^2 - h^2 + \kappa_m \kappa_\ell / 4)^2 + 2\kappa_m \kappa_\ell h (g + h)} \nonumber \\
    & = \bar{\alpha} \frac{1}{C\varepsilon^2/4 + 1}
\end{align}
where $\bar{\alpha} = \alpha_0(0)|_{\kappa_m = \kappa_\ell} = 2 \kappa_a n_a/\kappa_\ell$, and where in the second line we omit two terms from the denominator under the approximation of large $C$ and small $\varepsilon$. This expression shows that the on-resonance critically-coupled visibility is maintained when $|\varepsilon| \ll 2/\sqrt{C}$. For large $C$, then, the scan rate enhancement has much less restrictive scaling than the reflection susceptibility.

To qualitatively understand why the scan rate enhancement is much less sensitive than the reflection susceptibility to deviations from $g=h$, we can first consider $g > h$. The extreme sensitivity of $\chi_{mm}$ to slight mismatch in this regime is a consequence of destructive interference between two additive terms, an effect not relevant for the transmission susceptibilities $\chi_{ma}$ and $\chi_{m\ell}$ given by Eqs.\ \eqref{eq:chi_ma} \eqref{eq:chi_ml}. Thus, while increasing $g$ relative to $h$ has a dramatic effect on how unamplified measurement noise is routed through CEASEFIRE, the gain experienced by the axion signal and cavity noise rolls off more slowly. While increasing $g$ reduces the gain, it also increases the bandwidth, and as a result the scan rate enhancement is actually maximized for $g$ slightly larger than $h$, as shown in Fig.\ \ref{fig:ghmismatch}(b).

The formal insensitivity of the scan rate enhancement to $h > g$ deviations is a consequence of the fact that $\chi_{mm}$, $\chi_{ma}$, and $\chi_{m\ell}$ all exhibit the same divergent gain for $\varepsilon \rightarrow -1/2C$, and thus this divergence cancels in the expression for the visibility. It should be emphasized that this is merely an artifact of the formalism: in reality, the system would cross the parametric oscillation threshold at this point, and could not be operated to deliver a meaningful scan rate enhancement in the hatched region of Fig.\ \ref{fig:ghmismatch}(b).

\section{Extended CEASEFIRE model}
\label{appendix:transmissionline}

In this appendix, we describe in detail the calculations used to obtain the scan rate enhancement presented in Sec.\ \ref{sec:scanRate}. In the first subsection, we construct a lumped-element circuit model of the system shown in Fig.\ \ref{fig:schematic}(a) and derive the normal mode frequencies and loss rates. In Appx.\ \ref{subsection:interactionRates}, we derive expressions for intermode interaction rates in terms of  circuit model parameters. In Appx.\ \ref{subsection:IOtheory}, we present relevant susceptibility matrix elements for an input-output theory model that has been extended to include the effects of the two modes spectrally closest to the cavity mode. In Appx.\ \ref{subsection:SRE}, we discuss the scan rate enhancement predicted by this extended model.

\subsection{Circuit model and normal mode identification} 
\label{subsection:circuit_model}

\begin{figure*}[t]
	\centering
	\includegraphics[scale=1]{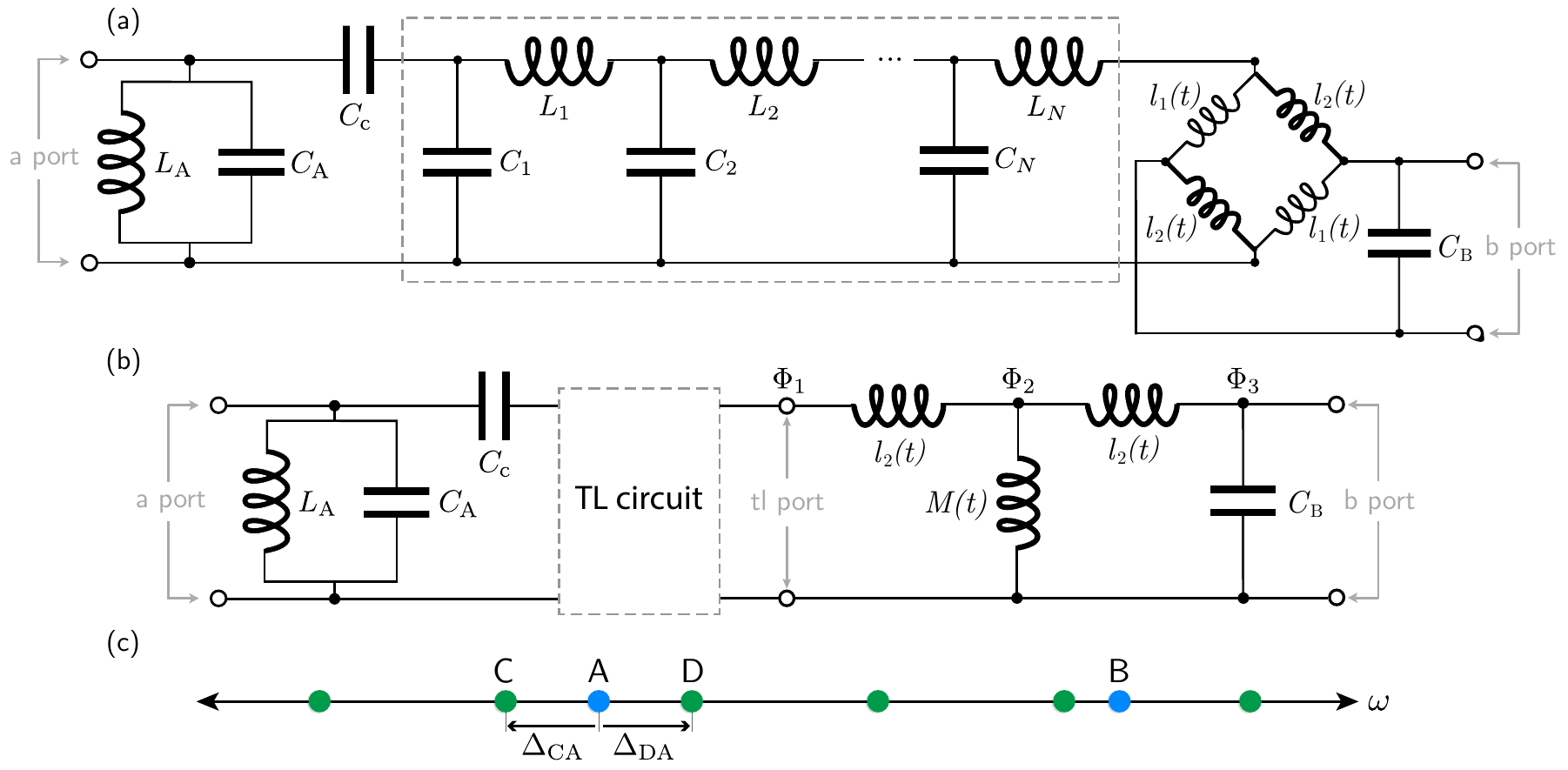}
	\caption{CEASEFIRE circuit model. (a) A coupling capacitor ($C\textsubscript{c}$) couples a parallel $LC$ resonator ($L\textsubscript{A}$ and $C\textsubscript{A}$), representing the cavity, to an inductor-capacitor ladder circuit ($L_i$ and $C_i$; $i=1,\dots,N$) representing the transmission line. The ladder circuit is connected to a readout resonator formed by a capacitor $C\textsubscript{B}$ in parallel with an inductive Wheatstone bridge. The inductance of each element in the bridge is composed of a static piece $L_0$ and a modulation $M(t)$; the sign of the modulation is opposite for adjacent elements, i.e. the time-varying inductance of one pair of opposing bridge inductors has time-varying inductance $l_1 = L_0 + M(t)$, while the other has $l_2 = L_0 - M(t)$. Ports \textsf{a} and \textsf{b} represent the cavity loss and readout resonator measurement port, respectively. (b) The transmission line circuit elements are encompassed in a dashed box representing a two-port impedance network; one of these ports, relevant to calculating couplings between the readout mode and transmission line modes, is labeled the \textsf{tl} port. The Wheatstone bridge can be equivalently represented by a T-junction of inductors $l_2$ and $M(t)$. The fluxes at each node, given by $\Phi_j = \int \text{d}t \,V_j$, are used as the normal coordinates of the system. (c) The normal modes of the circuit model comprise the cavity mode, labeled A, the readout mode, labeled B, and the discrete spectrum of evenly-spaced transmission line modes. The two transmission line modes spectrally nearest to the A mode, labeled the C and D modes, are separated from the A mode by $\Delta\textsubscript{CA}=\omega\textsubscript{C}-\omega\textsubscript{A}$ and $\Delta\textsubscript{DA}=\omega\textsubscript{D}-\omega\textsubscript{A}$.}
	\label{fig:TL_circuit_model}
\end{figure*}

A lumped-element circuit model allows us to  identify the normal modes and intermode couplings in a system comprising the cavity, readout resonator, and transmission line. We use the circuit model diagrammed in Fig.\ \ref{fig:TL_circuit_model}(a), in which the cavity is represented by a parallel $LC$ resonator (inductance $L\textsubscript{A}$ and capacitance $C\textsubscript{A}$) coupled to the transmission line by a coupling capacitor $C\textsubscript{c}$; the transmission line is modeled as an $LC$ ladder circuit with $N$ identical cells. Each cell in the ladder circuit introduces an additional mode to the system, and the total number of cells determines the cutoff in the mode spectrum of the transmission line. We include $N=400$ cells, resulting in predictions that are within 3\% of their asymptotic values.

The readout resonator comprises four time-varying inductors in a Wheatstone bridge configuration in parallel with a capacitor $C\textsubscript{B}$. In a practical implementation, the inductive elements in the bridge could be individual Josephson junctions whose inductance is modulated by differential current drives in the presence of a static external flux threading the bridge \cite{bergeal2010analog}, or SQUID arrays whose inductance is modulated by an external flux \cite{chapman2016general}. When the bridge is balanced (all inductors have equal inductance $L_0$), the potential between the north and south nodes is decoupled from the potential between the east and west nodes. External drives produce inductance modulation $\pm M(t)$ with $\langle M(t)\rangle = 0$ and opposite sign for adjacent inductors:
\begin{align}
    l_1(t)={}& L_0+M(t) \label{eq:definel1} \\
    l_2(t)={}& L_0-M(t), \label{eq:definel2}
\end{align}
where $l_1(t)$ $[l_2(t)]$ is the time-varying inductance of the northeast/southwest [northwest/southeast] inductors. This generates a dynamic, i.e.\ purely time-dependent, coupling between the readout circuit and the cavity and transmission line modes. 

To analyze the normal modes of this system, we reduce the Wheatstone bridge to an equivalent T-junction circuit [also equivalent to the mutual inductance representation used in Fig.\ \ref{fig:schematic}(a)], and group the transmission line circuit elements into a two-port impedance network [Fig.\ \ref{fig:TL_circuit_model}(b)]. One port of this network, labeled the \textsf{tl} port, will be used in the calculation of the interaction rates. We also introduce two other ports to the model: the readout circuit measurement port, labeled \textsf{b}, and a fictitious port modeling loss in the axion cavity, labeled \textsf{a}. We then calculate the admittance $Y_j(\omega)$, defined as the ratio of the short-circuit current that flows through port $j$ to the voltage imposed at the same port, when all other ports are left open. The normal mode frequencies ${\omega_i}$ are determined by the condition $\text{Im}[Y_{j}(\omega_i)]=0$; this prescription identifies the same normal modes regardless of which port $j$ is measured \cite{BBQ}.

To identify the normal modes that most resemble the uncoupled cavity and readout circuit modes, we define the effective impedance of the $i^\text{th}$ normal mode seen looking into port $j$ to be
\begin{equation}
    Z_{i,j}^\text{eff} = \frac{2}{\omega_i \operatorname{Im} Y_{j}'(\omega_i)}.
    \label{eq:Zeff}
\end{equation}
The modes with the maximum $Z_{i,\textsf{a}}^\text{eff}$ and the maximum $Z_{i,\textsf{b}}^\text{eff}$ (hereafter, the cavity and readout modes, $i=\text{A}$ and $i=\text{B}$ respectively) are the modes whose energy is most concentrated in the cavity and readout circuits, respectively.

Whatever fraction of the cavity mode energy is not spatially confined to the cavity does not contribute to axion-photon conversion, and thus dilution of the cavity mode energy degrades the scan rate. To quantify this effect we define the cavity mode self-participation $p\textsubscript{A}$ as the fraction of cavity mode energy that is localized in the $LC$ circuit representing the cavity, given by
\begin{equation}
    p\textsubscript{A} =\frac{Z_\text{A,\textsf{a}}^\text{eff}}{Z_\text{A}^{\text{ch}}},
    \label{eq:p_A}
\end{equation}
where $Z_\text{A}^{\text{ch}} = \sqrt{L\textsubscript{A}/C\textsubscript{A}}$ is the characteristic impedance of the cavity circuit. The self-participation may be increased by decreasing the coupling capacitance $C\textsubscript{c}$, though this will also reduce the dynamic coupling of the cavity mode to the readout and transmission line modes. In a physical implementation, the axion cavity external coupling is controlled by the insertion depth of an antenna that probes the cavity's electric field, and can be adjusted during a scan \cite{kenany2017}.

To incorporate loss into the circuit model, we introduce resistors $R\textsubscript{A}$ and $R\textsubscript{B}$ across the $\textsf{a}$ and $\textsf{b}$ ports respectively. The quality factor of the cavity mode is then given by
\begin{equation}
    Q\textsubscript{A} = \frac{\omega\textsubscript{A}}{2}\frac{\text{Im} Y_\text{AA}'(\omega_\text{A})}{\text{Re} Y_\text{AA}(\omega_\text{A})}, \label{eq:Qfactor}
\end{equation}
with the readout mode quality factor  $Q\textsubscript{B}$ defined analogously, and the loss rates are given by
\begin{equation}
    \kappa\textsubscript{A} = \frac{\omega\textsubscript{A}}{Q\textsubscript{A}}, \; \kappa\textsubscript{B}=\frac{\omega\textsubscript{B}}{Q\textsubscript{B}}.
    \label{eq:kappas}
\end{equation}

When the readout circuit is totally decoupled from the transmission line ($M(t)=0$), we denote the cavity and readout mode losses as $\kappa_{\text{A}0}$ and $\kappa_{\text{B}0}$ respectively. The uncoupled readout mode loss is given simply by $\kappa_{\text{B}0} = 1/(R\textsubscript{B}C\textsubscript{B})$; $\kappa_{\text{A}0}$ is likewise determined by $R\textsubscript{A}$ but no such simple expression exists because setting $M(t)=0$ does not decouple the cavity circuit from the transmission line.
We choose $R\textsubscript{A}$ such that $\kappa_{\text{A}0}/(2\pi) = 100$~kHz. This choice models a system in which the uncoupled cavity and transmission line modes have equal 100~kHz loss rates, and thus coupling the cavity to the transmission line does not change the cavity mode loss. In a practical implementation, such low losses for the standing wave modes could be achieved by separating the cavity from the readout circuit through a waveguide instead of a transmission line. An additional advantage of a waveguide-based implementation is a larger free spectral range. A waveguide of length $L = 50$~cm has a free spectral range $\Delta \textsubscript{FSR}/2\pi = 300$~MHz, and we assume this value throughout our analysis.

As we will show in the following section, turning on the inductance modulation generates the desired dynamic couplings but also induces static coupling between the readout mode and the cavity and transmission line modes; this static coupling will be suppressed but not completely eliminated by the Wheatstone bridge geometry. The partial hybridization between the cavity and readout modes causes $\kappa\textsubscript{A}$ and $\kappa\textsubscript{B}$ to deviate from $\kappa\textsubscript{A0}$ and $\kappa\textsubscript{B0}$ respectively. For a given modulation amplitude, we can relate the loss rates defined above to $\kappa_\ell$ and $\kappa_m$ used in the main text. We set $\kappa_\ell=\kappa\textsubscript{A}$, so that loss that the cavity mode inherits from the readout mode is taken into account in calculating the scan rate enhancement, and set $\kappa_m = \kappa_{\text{B}0}$, so that $R\textsubscript{B}$ models the external coupling of the readout mode rather than its total loss rate. Loss inherited by the readout mode is very small compared to $\kappa_m$ and thus has a negligible effect on the behavior of CEASEFIRE; we do not include it in our input-output theory models.

\subsection{Derivation of interaction rates}
\label{subsection:interactionRates}
Having identified the normal modes of the system, we now derive the dynamic coupling rates between normal modes. For any given inductance modulation $M(t)$, these dynamic coupling rates can be expressed entirely in terms of circuit parameters, and then used in the input-output theory framework of Sec.\ \ref{sec:inputOutput} to determinate the scan rate enhancement of CEASEFIRE.

We begin with the classical Hamiltonian $H_\text{T}$ of the T-junction circuit in Fig.\ \ref{fig:TL_circuit_model}(b),
taking as our normal coordinates the fluxes $\Phi_j=\int V_j \dif t$ at the nodes of the circuit. This Hamiltonian is
\begin{align}
     H\textsubscript{T} &= \frac{(\Phi_1-\Phi_2)^{2}}{2\,l_2(t)}+\frac{\Phi_2^{2}}{2\,M(t)}+\frac{(\Phi_2-\Phi_3)^{2}}{2\,l_2(t)} \nonumber \\ 
     &= \frac{L_0\left(\Phi_1^2 + \Phi_3^2\right)-2M(t)\Phi_1\Phi_3}{2l_1(t)l_2(t)} \label{eq:Tjunction},  
\end{align}
where $\Phi_1$, $\Phi_2$, and $\Phi_3$ are the fluxes at the left, center, and right nodes of the T-junction, respectively, and in the second line we have used Kirchoff's current law to eliminate $\Phi_2$. Note that for $\langle M(t) \rangle = 0$ there is still a $\Phi_1\Phi_3$ cross-term in $\langle H\textsubscript{T}\rangle$, where angle brackets denote a time average, due to the presence of time-dependence in the denominator; this is the origin of the residual hybridization of the cavity and readout modes noted above.

Following the procedure outlined in the preceding section to obtain the normal mode frequencies, effective impedances, and loss rates with $M(t)\neq0$ fully accounts for the  effects of the static interaction Hamiltonian $\langle H\textsubscript{T}\rangle$. Thus we can  identify the dynamic interaction Hamiltonian as
\begin{equation}
\label{eq:H_int}
\hat{H}\textsubscript{int} =   \hat{H}_{\text{BS}} + \hat{H}_{\text{TMS}} = \hat{H}\textsubscript{T} - \langle \hat{H}\textsubscript{T}\rangle,
\end{equation}
where we have promoted the fluxes to quantum operators.

The normal coordinates $\hat{\Phi}_j$ can generally be expressed as linear combinations of the normal mode field operators $\hat{a}_i$ and $\hat{a}_i^\dag$, with coefficients given by the effective impedances defined in Eq.\ \eqref{eq:Zeff}. In particular, the fluxes $\hat{\Phi}_1$ and $\hat{\Phi}_3$ are defined at the nodes where we defined the \textsf{tl} and \textsf{b} ports in Appx.\ \ref{subsection:circuit_model}; their normal mode expansions thus take the form 
\begin{equation}
    \hat{\Phi}_j=\sum_{i}{\sqrt{\frac{Z_{i,j}^\text{eff}}{2}} (\hat{a}_i+\hat{a}_i^{\dagger})},
    \label{eq:fluxExpansion}
\end{equation}
where $Z_{i,1}^{\text{eff}}=Z_{i,\textsf{tl}}^{\text{eff}}$ and $Z_{i,3}^{\text{eff}}=Z_{i,\textsf{b}}^{\text{eff}}$. Elsewhere in the text, the normal mode field operators are denoted $\hat{A} = \hat{a}_\text{A}$, $\hat{A}^\dag = \hat{a}_\text{A}^\dag$, $\hat{B} = \hat{a}_\text{B}$, and so on.

We now assume that the fractional inductance modulation $\epsilon(t) = M(t)/L_0$ is small, and given by
\begin{equation}
    \epsilon(t) = \ell_\Delta\cos(\omega_\Delta t)+\ell_\Sigma\cos(\omega_\Sigma t+\phi),
\end{equation}
where $\ell_\Delta$ and $\ell_\Sigma$ are set by the amplitudes of the difference- and sum-frequency drives. Expanding the interaction Hamiltonian to first order in $\epsilon(t)$ yields
\begin{equation}
     \hat{H}\textsubscript{int} \approx -\frac{1}{L_0} \epsilon (t) \hat{\Phi}_1 \hat{\Phi}_3.
     \label{eq:interaction}
\end{equation}
The normal mode expansions of $\hat{\Phi}_1$ and $\hat{\Phi}_3$ yield $\hat{A}\hat{B}^{\dagger}$ and $\hat{A}^{\dagger}\hat{B}^{\dagger}$ cross terms; as in Eq.\ \eqref{eq:interactionHamiltonian}, the coefficients of these terms give the state swapping and two-mode squeezing interaction rates between the cavity mode and readout mode. The same prescription can be used to identify interaction rates between other pairs of modes. To first order in $\epsilon(t)$, the symmetry of the Wheatstone bridge dictates that the normal modes separate into two classes: the cavity and transmission line modes contribute only to the flux at the left node of the T-junction ($Z_{i,\textsf{b}}^{\text{eff}}=0$ for these modes), and the readout mode contributes only to the flux at the right node ($Z_{B,\textsf{tl}}^{\text{eff}}=0$). 
The interaction rates between a mode $i$ and the readout mode are thus given by
\begin{align}
    g_{i\text{B}} ={}& \frac{l_\Delta \sqrt{Z_{i,\textsf{tl}}^\text{eff} Z_{B,\textsf{b}}^\text{eff}}}{4 L_0}
    \label{eq:g_ij},\\[2mm]
    h_{i\text{B}} ={}& \frac{l_\Sigma \sqrt{Z_{i,\textsf{tl}}^\text{eff}Z_{B,\textsf{b}}^\text{eff}}}{4 L_0}.
    \label{eq:h_ij}
\end{align}
For any mode $i$, the interaction rates $g_{i\text{B}}$ and $h_{i\text{B}}$ are matched when the difference-frequency and sum-frequency drives produce equal fractional inductance modulation.

In the following subsection, we derive the scan rate enhancement in the presence of transmission line modes to first order in $\epsilon(t)$. The results presented in Fig.\ \ref{fig:visibility} also account for three effects that arise at higher orders in $\epsilon(t)$. First, as noted above, turning on the modulation causes the cavity mode to inherit loss from the readout mode; the first nonzero contribution to this inherited loss enters at second order in $\epsilon(t)$.  Second, the static coupling induced by the modulation alters the effective normal mode impedances $Z_{i,\textsf{tl}}^{\text{eff}}$ and $Z_{B,\textsf{b}}^{\text{eff}}$  and thus modifies the dynamic coupling rates. Finally, we account for the effects of new couplings that appear when Eq.\ \eqref{eq:H_int} is expanded to higher order in $\epsilon(t)$:
\begin{align}
     \hat{H}\textsubscript{int} &= \frac{1}{L_0}\left[ \left(\epsilon^2 + \epsilon^4 + \dots\right)\left(\hat{\Phi}_1^2 + \hat{\Phi}_3^2\right)\right. \nonumber\\ &- \left.\left(\epsilon + \epsilon^3 + \dots\right) \hat{\Phi}_1 \hat{\Phi}_3\right] \nonumber\\ &- \frac{1}{L_0}\left[ \left(\langle\epsilon^2\rangle + \langle\epsilon^4\rangle + \dots\right)\left(\hat{\Phi}_1^2 + \hat{\Phi}_3^2\right)\right. \nonumber\\ &- \left.\left( \langle\epsilon^3\rangle + \dots\right) \hat{\Phi}_1 \hat{\Phi}_3\right].
     \label{eq:interaction_higher}
\end{align}

The $\epsilon^2\hat{\Phi}_1^2$ and $\epsilon^2\hat{\Phi}_3^2$ terms generate single-mode squeezing interaction for the cavity and readout modes, respectively. When the single-mode squeezing rates exceed the decay rates of these modes, the visibility on resonance is suppressed and thus the scan rate enhancement is substantially reduced. We can compensate for these undesired interactions by introducing additional drives at $2\omega_\text{A}$ and $2\omega_\text{B}$. We numerically optimize scan rate with respect to the amplitudes of these additional drives to obtain the results presented in Fig.\ \ref{fig:visibility}.

\subsection{Four-mode input-output theory}
\label{subsection:IOtheory}

In Sec.\ \ref{sec:transmissionLine} we asserted that the transmission line modes with the most impact on the scan rate enhancement are those spectrally closest to the cavity mode, labeled C and D in Fig.\ \ref{fig:TL_circuit_model}(c). Here we quantify the effects of these modes on the scan rate enhancement using a four-mode input-output theory model that also includes two additional ports representing the C and D mode internal losses and the noise incident on these ports. In the main text, we defined $\omega_\Delta=\omega\textsubscript{B} - \omega\textsubscript{A}$; for other pairs of modes we will use the notation $\Delta_{ij} = \omega_i - \omega_j$. We follow the procedure described in Appx.\ \ref{appendix:inputoutput} to derive relevant elements of the four-mode susceptibility matrix. In the rotating frame of the readout mode, the equations of motion are given by
\begin{align}
\vspace{2mm}
\begin{split}
    \dot{\hat{A}} ={}& \bigg(i \omega_\Delta -\frac{\kappa_\ell}{2}\bigg) \hat{A}+\sqrt{\kappa_a} \mathop{\hat{\xi}_{\text{in},a}} + \sqrt{\kappa_\ell} \mathop{\hat{\xi}_{\text{in},\ell}} \\
        & - i g\textsubscript{AB} \mathop{\hat{B}} e^{i \omega_\Delta t} - i h\textsubscript{AB} \, e^{-i \phi}\mathop{\hat{B}^{\dagger}} e^{i \omega_\Delta t}
\end{split}\\[2mm]
\begin{split}
    \dot{\hat{B}} = {}& -\frac{\kappa_m}{2}\mathop{\hat{B}} + \sqrt{\kappa_m}\mathop{\hat{\xi}_{\text{in},m}} - i g\textsubscript{AB} \mathop{\hat{A}} e^{-i\omega_\Delta t} \\
        & -i h\textsubscript{AB} \, e^{-i\phi} \hat{A}^{\dagger}e^{i \omega_\Delta t} -i g\textsubscript{CB} \mathop{\hat{C}} e^{-i \omega_\Delta t} \\
        & -i h\textsubscript{CB} \, e^{-i\phi} \mathop{\hat{C}^{\dagger}} \, e^{i \omega_\Delta t}-i g\textsubscript{DB} \mathop{\hat{D}} e^{-i \omega_\Delta t}\\
        & -i h\textsubscript{DB} \, e^{-i \phi} \mathop{\hat{D}^{\dagger}} e^{i \omega_\Delta t}
\end{split}\\[2mm]
\begin{split}
    \dot{\hat{C}} ={}& \bigg(i\Delta\textsubscript{BC}-\frac{\kappa\textsubscript{C}}{2}\bigg)\mathop{\hat{C}}+\sqrt{\kappa\textsubscript{C}}\mathop{\hat{\xi}_{\text{in,C}}} -i g\textsubscript{CB} \mathop{\hat{B}} e^{i \omega_\Delta t}\\
    & - i h\textsubscript{CB} \, e^{-i\phi} \mathop{\hat{B}^{\dagger}} e^{i \omega_\Delta t}
\end{split}\\[2mm]
\begin{split}
    \dot{\hat{D}} ={}&
    \bigg(i\Delta\textsubscript{BD}-\frac{\kappa\textsubscript{D}}{2}\bigg)\hat{D}+\sqrt{\kappa\textsubscript{D}}\mathop{\hat{\xi}_{\text{in,D}}} - i g\textsubscript{DB} \mathop{\hat{B}} e^{i \omega_\Delta t} \\
    & -i h\textsubscript{DB} \, e^{-i\phi}\mathop{\hat{B}^{\dagger}} e^{i \omega_\Delta t} .\raisetag{5mm}
\end{split}
\end{align}
We solve these equations in the frequency domain and apply the general input-output relations given by Eq.\ \eqref{eq:inputOutputRelations}, extended to include the two additional ports, to obtain susceptibility matrix elements. The susceptibilities relevant to the measurement port output field are
\begin{align}
    \begin{split}
        \zeta^{(4)}_{mk}(\omega)={}& i\,\eta^{-1}(\omega)[\,g\textsubscript{AB}\,\beta^* (-\omega)\\
        & \hspace{1.5cm} -h\textsubscript{AB} \, e^{i\phi} \, \gamma(\omega)] \sqrt{\kappa_m\kappa_k}
        \label{eq:chimk}
    \end{split}\\[2mm]
    \begin{split}
        \zeta^{(4)}_{m k^\dag}(\omega)={}& i\,\eta^{-1}(\omega)[\,h\textsubscript{AB}\,e^{-i\phi} \, \beta^*(-\omega) \\
        & \hspace{1.5cm} -g\textsubscript{AB} \, \gamma(\omega)] \sqrt{\kappa_m\kappa_k}
    \end{split}\\[2mm]
        \begin{split}
        \zeta^{(4)}_{mj}(\omega)={}& i\,\eta^{-1}(\omega) [\, g_{j\text{B}}\, \beta^*(-\omega)-h_{j\text{B}} \, e^{i\phi} \, \gamma (\omega)]\\ & \hspace{1.5cm} \times\left[\frac{\sqrt{\kappa_j\kappa_m}\left(i\omega+\frac{\kappa_\ell}{2}\right)}{i\left(\omega+\Delta_{j\text{A}}\right)+\frac{\kappa_j}{2}}\right]
    \end{split}\\[2mm]
    \begin{split}
        \zeta^{(4)}_{mj^\dag}(\omega) ={}& 
        i \, \eta^{-1}(\omega)[ \, h_{j\text{B}}\, e^{-i\phi} \beta^\ast(-\omega) -g_{j\text{B}} \, \gamma (\omega)]\\
        & \hspace{1.5cm} \times\left[ \frac{\sqrt{\kappa_j \kappa_m} \left( i \omega + \frac{\kappa_\ell}{2} \right) }{i(\omega-\Delta_{j\text{A}})+\frac{\kappa_j}{2}}\right] \label{eq:chimjdag}
    \end{split}\\[2mm]
    \zeta^{(4)}_{mm}(\omega)={}& 1-\eta^{-1}(\omega)\,\beta^* (-\omega)\,\kappa_m \! \left(i\omega+\frac{\kappa_\ell}{2}\right)
    \label{eq:chimm}
    \\[2mm]
    \zeta^{(4)}_{mm^{\dagger}}(\omega)={}& - \! \eta^{-1}(\omega) \, \gamma(\omega) \, \kappa_m \! \left(i\omega+\frac{\kappa_\ell}{2}\right)
    \label{eq:chimmdag}
\end{align}
as well as their conjugates, where $k = \{ a, \ell \}$, $j =\{ \text{C},\, \text{D} \}$, and we have defined
\begin{align}
    \eta(\omega)={}& \, \beta (\omega) \beta^* (-\omega)-\gamma (\omega)\gamma^*(-\omega),\\[3mm]
    \begin{split}
        \beta(\omega)={}& \left(i\omega + \frac{\kappa_m}{2}\right) \left(i\omega + \frac{\kappa_\ell}{2} \right) + \left(g_{\text{AB}}^2 - h_{\text{AB}}^2\right)\\
        & + \sum_{j} \left[ \frac{g_{j\text{B}}^2}{i (\omega + \Delta_{j\text{A}}) + \frac{\kappa_j}{2}}-\frac{h_{j\text{B}}^2}{i (\omega-\Delta_{j\text{A}}) + \frac{\kappa_j}{2}} \right]\\[.5mm]
        & \hspace{13mm} \times\left(i\omega+\frac{\kappa_\ell}{2}\right), \raisetag{2.67cm} \label{eq:beta}
    \end{split} \\[3mm]
    \gamma (\omega) ={}&
    \sum_j \frac{2i \, g_{j\text{B}} \, h_{j\text{B}}\,e^{-i\phi} \left(i\omega + \frac{\kappa_\ell}{2} \right) \Delta_{j\text{A}}} {\left[ i (\omega - \Delta_{j\text{A}}) + \frac{\kappa_j}{2} \right] \left[ i (\omega+\Delta_{j\text{A}})+\frac{\kappa_j}{2}\right]}. \raisetag{1.12cm}
\end{align}
These expressions indicate that the presence of transmission line modes affects the behavior of CEASEFIRE in three ways: the susceptibilities governing the transmission of fields from the ports on the axion cavity to the measurement port are modified, noise from the internal loss of the transmission line modes can be transmitted to the measurement port, and the measurement port reflection susceptibility is also modified.

We now consider these three effects in turn, in each case assuming $g\textsubscript{AB} = h\textsubscript{AB}$, $g_{j\text{B}} = h_{j\text{B}}$, and $\kappa\textsubscript{A} = \kappa_j \ll \Delta_{j\text{A}}$ for both $j=\text{C}$ and $j=\text{D}$. Working out the axion mode transmission susceptibilities subject to these conditions, we find that
\begin{equation}
\frac{\zeta_{mk}^{(4)}(\omega)}{\zeta_{mk}(\omega)} = \frac{\zeta_{mk^\dag}^{(4)}(\omega)}{\zeta_{mk^\dag}(\omega)} = 1.
\end{equation}
That is, when interaction rates are matched, the CEASEFIRE transmission susceptibilities are unchanged from their values in the two mode model.

Next, we evaluate the gain experienced by noise from the $j$ mode internal loss ($j = \text{C},\text{D}$) to the gain experienced by noise from the cavity mode internal loss; to qualitatively illustrate the behavior, it is sufficient to consider the gain on resonance. This gain ratio is
\begin{equation}
    \left|\frac{\zeta_{mj}^{(4)}(0)}{\zeta_{m\ell}^{(4)}(0)}\right|^2 = \left(\frac{g_{j\text{B}}}{g_\text{AB}}\right)^2\frac{\kappa_\ell^2}{4\Delta_{j\text{A}}^2}.
\end{equation}
With the circuit parameters that optimize the CEASEFIRE scan rate, $g_{j\text{B}} \approx 3g_{\text{AB}}$, because the coupling capacitor $C\textsubscript{c}$ reduces the contribution of the cavity mode to the flux at the \textsf{tl} port relative to the contribution from the transmission line modes. However, $\kappa_\ell \sim \left(10^{-2 } - 10^{-3}\right)\Delta_{j\text{A}}$ in roughly 95\% of the cavity mode tuning range. In other words, the large detuning of the transmission line modes from the cavity mode suppresses transmission of amplified noise from the internal loss of these modes, and the scan rate enhancement is preserved.

Finally, we consider the change in the measurement port reflection susceptibility, which turns out to be the most significant of these effects. Again it is sufficient to consider the behavior of the susceptibility on resonance. The phase-preserving reflection susceptibility becomes
\begin{equation}
\zeta^{(4)}_{mm}(0) = -1 + \frac{8i}{\kappa_m}\left(\frac{g_{\text{CB}^2}}{\Delta_{\text{CA}}} + \frac{g_{\text{DB}^2}}{\Delta_{\text{DA}}}\right),
\label{eq:chimmfourmode}
\end{equation}
and the phase-conjugating reflection susceptibility becomes
\begin{equation}
\zeta^{(4)}_{mm^\dag}(0) = -\frac{8i}{\kappa_m}\left(\frac{g_{\text{CB}^2}}{\Delta_{\text{CA}}} + \frac{g_{\text{DB}^2}}{\Delta_{\text{DA}}}\right)e^{-i\phi}.
\label{eq:chimmdagfourmode}
\end{equation}
These susceptibilities do not have unit magnitude, and thus measurement noise incident on the measurement port can be amplified in reflection (a similar effect occurs for an $h>g$ mismatch; see Appx.\ \ref{appendix:ghMismatch}).

The amplification of reflected measurement noise can be traced to the third term of Eq.\ \eqref{eq:beta}, in which $\Delta_{j\text{A}}$ appears with the opposite sign in the $g_{j\text{B}}^2$ and $h_{j\text{B}}^2$ terms, such that these terms do not cancel when $g_{j\text{B}} = h_{j\text{B}}$. This in turn is a result of the fact that the detuning of the difference-frequency drive the from the B mode/$j$ mode frequency difference and the detuning of the sum-frequency drive from the B mode/$j$ mode frequency sum have opposite sign. Assuming the hierarchy of mode frequencies illustrated in Fig.\ \ref{fig:TL_circuit_model}(c) for concreteness, we see that $\omega_\Delta$ is larger than $\Delta\textsubscript{BD}$ by $\Delta\textsubscript{DA}$ while $\omega_\Sigma$ is smaller than $\omega\textsubscript{B} + \omega\textsubscript{D}$ by $\Delta\textsubscript{DA}$. The situation is reversed for the C mode: $\omega_\Delta < \Delta\textsubscript{BC}$ and  $\omega_\Sigma > \omega\textsubscript{B} + \omega\textsubscript{C}$.

However, when the A mode is spectrally centered between and C and D modes ($\Delta\textsubscript{CA} = -\Delta\textsubscript{DA}$), the D mode compensates for the imbalance of the drive detunings as seen by the C mode and vice versa. If we further assume that  $g\textsubscript{CB}=g\textsubscript{DB}$ (generically true to a good approximation, since spectrally close transmission line modes contribute comparably to the flux at the \textsf{tl} port), the contributions of the C and D modes to the reflection susceptibilities Eqs.\ \eqref{eq:chimmfourmode} and \eqref{eq:chimmdagfourmode} cancel, and we recover the two-mode behavior. All effects of transmission line modes further detuned from the A mode are suppressed by the large detuning, and exhibit the same cancellation when the A mode is spectrally centered between the C and D modes.
\begin{figure*}[!htp]
	\centering
	\includegraphics[scale=1]{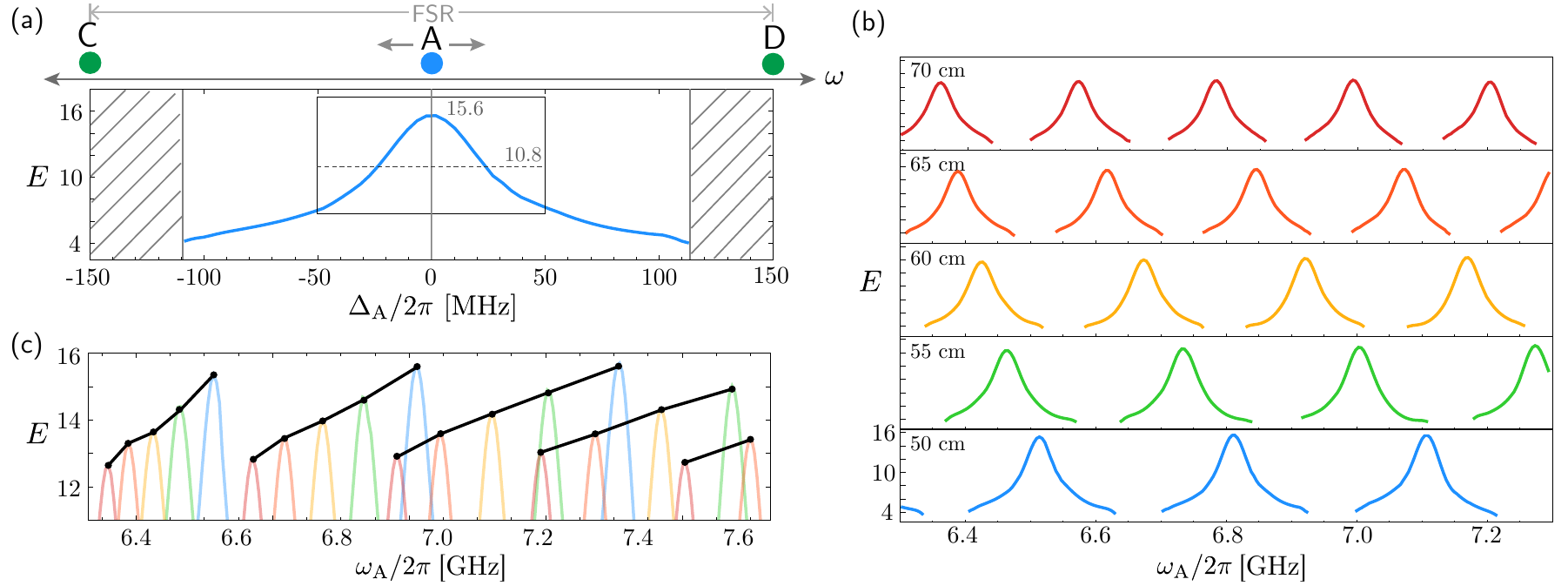}
	\caption{Scan rate enhancement with cavity mode tuned between transmission line modes, using the same parameters as in Fig.\ \ref{fig:visibility}(a). Panel (a) shows the scan rate enhancement over one FSR, assuming a 50 cm line such that $\Delta \textsubscript{FSR}/2\pi = 300$ MHz. The scan rate enhancement peaks when the cavity mode is midway between the C and D modes. Away from this point, the scan rate enhancement decreases due to measurement noise being amplified in reflection; the average scan rate enhancement within a 100 MHz window centered at $(\omega\textsubscript{C} + \omega\textsubscript{D})/2$ is 10.8. As the cavity mode nears a transmission line mode, the modes have an avoided crossing, in which no normal mode solution exists. These inoperable frequency ranges are shown by hatched regions. Panel (b) shows the scan rate enhancement for five different transmission line lengths ranging from 50 to 70 cm. The transmission line length sets the free spectral range, altering the frequencies at which the scan rate enhancement peaks. Panel (c) shows the scan rate enhancement with a transmission line of continuously tunable length. Black lines connect the peaks between various transmission line lengths to indicate the scan rate enhancement achievable by tuning the length continuously.}
	\label{fig:TL length}
\end{figure*}

\subsection{Scan rate enhancement and transmission line length variation}
\label{subsection:SRE}

We find the scan rate enhancement in the extended model following the same procedure as in Appx.\ \ref{appendix:inputoutput}. The definition of the four-mode input spectral density matrix $\textbf{S}^{(4)}_\text{in}$ is analogous to Eq.\ \eqref{eq:inputSpectralDensityMatrix}, with additional terms for the vacuum and thermal noise at the C and D ports, and and the output spectral density matrix is given by
\begin{equation}
    \textbf{S}^{(4)}_{\text{out}} (\omega) = \left[ \boldsymbol{\zeta}^{\, (4)}(\omega)\right]^* \textbf{S}^{(4)}_\text{in} \left[\boldsymbol{\zeta}^{\, (4)}(\omega) \right] ^\text{T}.
\end{equation}
To calculate the measurement port output spectral density we require only the susceptibility matrix elements given by Eqs.\ \eqref{eq:chimk} through \eqref{eq:chimmdag} and their conjugates.
We then transform the measurement port output spectral density to the quadrature basis, and define the visibility $\alpha^{(4)}_\text{CF}(\omega)$ in terms of amplified-quadrature output spectral densities in the amplified quadrature. The scan rate enhancement in the extended model is then given by
\begin{equation}
    E^{(4)} = \frac{\int \dif \omega \, [\alpha^{(4)}_\text{CF}(\omega)]^2p_\text{A}^2}{\int \dif \omega \, [\alpha_0 (\omega)]^2_{\kappa_m = 2\kappa_\ell}},
\end{equation}
where $p\textsubscript{A}$ is the cavity mode self-participation defined in Eq.\ \eqref{eq:p_A}. To simplify the notation, we denote the four-mode scan rate enhancement in Figs.\ \ref{fig:visibility} and \ref{fig:TL length} simply as $E$. 

In Fig.\ \ref{fig:visibility} in the main text we plotted the scan rate enhancement for the special case $\omega\textsubscript{A} = (\omega\textsubscript{C} + \omega\textsubscript{D})/2$. Fig.\ \ref{fig:TL length}(a) shows the scan rate enhancement over one free spectral range as a function of cavity mode detuning $\Delta\textsubscript{A}$ from this optimal operating frequency, using the same parameters as in Fig.\ \ref{fig:visibility}(a). The peak scan rate enhancement midway between transmission line modes is $E = 15.6$ and the average scan rate enhancement over a 100 MHz range centered at this point is $\bar{E}=10.8$. To scan continuous regions of axion parameter space over more than one free spectral range, the transmission line modes can be shifted in frequency by adjusting the length of the line. Fig.\ \ref{fig:TL length}(b) shows the scan rate enhancement as a function of frequency for five transmission line lengths ranging from 50 to 70 cm. Continuously tuning the transmission line would allow for a scan rate enhancement near its maximum value over frequency ranges much larger than one free spectral range, as shown in Fig.\ \ref{fig:TL length}(c).

\clearpage

\bibliography{main}

\end{document}